\documentclass{aa}

\usepackage{graphicx}
	\graphicspath{{./img/},{./}}
\usepackage{txfonts}
\usepackage{hyperref}
\usepackage{subcaption}
\usepackage{xcolor}
\usepackage{textcomp,mathabx}
\usepackage{siunitx}
    \DeclareSIUnit{\astronomicalunit}{AU}
	\DeclareSIUnit{\parsec}{pc}
	\DeclareSIUnit{\earthmass}{M_\Earth}
	\DeclareSIUnit{\solarmass}{M_\Sun}
	\DeclareSIUnit{\jupitermass}{M_J}
	\DeclareSIUnit{\year}{yr}
    \DeclareSIUnit{\jansky}{Jy}
    \DeclareSIUnit{\arcsecond}{arcsec}
\usepackage{placeins}

\begin{document} 

    \bibliographystyle{aa}
    \title{Spiral formation caused by late infall onto protoplanetary disks}

    \author{L.-A. H\"uhn\inst{1}
        \and
        C. N. Kimmig\inst{2}
        \and
        C. P. Dullemond\inst{1}
        }

    \institute{Institut für Theoretische Astrophysik, Zentrum für Astronomie der Universität Heidelberg, Albert-Ueberle-Str. 2, 69120 Heidelberg, Germany\\
    \email{\href{mailto:huehn@uni-heidelberg.de}{huehn@uni-heidelberg.de}}
    \and
    Dipartimento di Fisica, Università degli Studi di Milano, Via Giovanni Celoria 16, 20133 Milano, Italy
    }

    \date{\today}
 
    \abstract
    {The classical picture that planet formation occurs in protoplanetary disks that are isolated from their environment is undergoing a major shift toward a more connected picture. An increasing amount of evolved disks are found to be actively interacting with their environment, often showing various types of spiral structures. In this work, we aim to investigate if these spirals can be a direct result of ongoing late infall using the grid-based 3D hydrodynamics code \texttt{FARGO3D}. We perform a detailed analysis of the spiral properties and appearance in scattered light and CO line emission using radiative transfer modeling with \texttt{RADMC3D}. In scattered light, we find both well-defined spirals with few arms ($m=2$) and more flocculent structures: The gradual accretion of gas remnants after a major accretion event has the most success in the former, whereas active accretion via streamers favors the latter. The $m=2$ spirals we find have a very low pattern speed, making them easily discernible from spirals caused by a perturber. We also find spiral patterns in the $^{12}$CO residual motions, but their morphology does not match the one found in scattered light. The disk perturbations are strongest in the upper layers ($z>4H$), which is reflected by the reduced amplitude of the residual motions in the more optically thin $^{13}$CO emission. Moreover, we find that the formation of $m=2$ spirals is not promoted in disks with lower mass, despite being more susceptible to deeper kinematic perturbations. While the late-infall streamers impact planet formation directly through the delivery of fresh material, we show that the midplane remains unperturbed unless the infalling mass is of the same order of magnitude as the disk mass. Planet formation can therefore only be impacted by late infall through secondary mechanisms that lead to dust trapping or the generation of turbulence starting from surface-level perturbations.}


    \maketitle
\section{Introduction}
In recent years, significant revelations have been discovered while trying to answer the question of how exoplanets form (for a review, see \citealt{drazkowska2023}), from both the  theoretical and observational perspectives. One of the big challenges in the field is understanding the coagulation of initially small dust grains in protoplanetary disks, the birthplaces of planets. As they grow, despite many still unexplained challenges, these grains form the necessary building blocks for planets. Observed substructures such as rings, gaps, and spirals in these disks can significantly influence and shape the growth process (for a review, see \citealt{bae2023}). Within the last few decades, many questions have arisen that are yet to be fully answered, concerning, among others, the available dust budget, its distribution through the disk, the timing of the emergence of the substructures, and whether they cause planet formation or are a consequence thereof.

Planet formation theory has classically treated the Class~II evolutionary stage of a disk, when most of the material from the natal cloud has disappeared, as the starting point. Consequently, planet formation has often been modeled as isolated from the disk's environment. However, recent observations have challenged this picture (e.g., \citealt{ginski2021,huang2021,garufi2022,gupta2024,speedie2024}), resulting in a large paradigm shift, where models specifically target the environmental interactions and their impact on the disk. Especially, the ongoing interaction of early-stage disks with their natal envelope has been the focus of recent work (e.g., \citealt{drazkowska2018,morbidelli2022,huehn2025a,carrera2025}), aiming to solve the issue of insufficient solid mass being available to form known giant planets during Class~II (\citealt{manara2018}, though it is debated, \citealt{savvidou2025}) by invoking that the formation of planetesimals, the precursors of planets, could start early, when more solids are available.

Observations indicate a ubiquitous amount of disks that are actively interacting with their environment also at later stages in their evolution, evident by large-scale structures, such as arms and tails (commonly refereed to as streamers), that are suggested to result from infall. In fact, ${\sim}30\%$ of observed disks in the Taurus star forming region exhibit these kinds of structures \citep{garufi2024}. Thus, it is vital to investigate the nature of these large-scale structures and how they influence the formation of planets. Similar to early-stage infall, that can provide a large amount of solids \citep{huehn2025a}, late accretion could deliver new, potentially chemically distinct material to the disk, influence its kinematics, and introduce substructure that could aid, or harm, the planet formation process. It could even form a new disk altogether \citep{kuffmeier2020,huehn2025b}.

Streamers can be interpreted as material that is accreted sporadically from the environment (also called late infall). Simulations with simplified models of the capture of a spherical gas cloudlet by a star and disk system \citep{dullemond2019} have facilitated the characterization of the structures seen in AB Aur \citep{calcino2025b} and DG Tau \citep{hanawa2024} as streamers. However, such cloudlet encounters might be rare, and these models may be an oversimplification; in reality, it might be necessary to assume that late infall is a result of Bondi-Hoyle-Lyttleton (BHL) accretion in the turbulent interstellar medium (ISM), for instance, from molecular remnants of the parent giant molecular cloud. Models of this kind of interaction are able to produce direct \citep{hd25} and indirect \citep{winter2024} infall characteristics of disks.

Next to the direct evidence provided by streamers, an indirect indicator of interaction of the disk with the environment are warps. With the increasing observational indications of warps and misaligned inner disks \citep{ansdell2020,bohn2022,villenave2024,kimmig2025,winter2025}, it is important to understand their origin. Late infall is one of the prime candidates for the creation of these warps \citep{kuffmeier2021,dullemond2022}, further indicating that infall might be common also at later disk stages. Other scenarios to form a warp include gravitational torques from stellar flybys (e.g., \citealt{cuello2023,kimmig2026}), inclined binaries \citep{facchini2013}, or inclined planets \citep{Nealon2018,zhu2019}. A warped disk can naturally exhibit additional substructure like spirals if it is also twisted \citep{winter2025}, or shadows \citep{marino2015,facchini2018}, which can, in turn, lead to the formation of spirals (e.g., \citealt{su2024,zhang2024,zhu2025,ziampras2025}).

Several systems with streamers and ongoing infall exhibit spiral structures. The spiral pattern can either consist of many flocculent arms like in SU~Aur \citep{ginski2021} or of a few, well-defined ones like in DR~Tau \citep{mesa2022,garufi2024}. Previous work investigating the formation of these structures revealed that infall could be the direct cause of $m=1$ spirals \citep{hennebelle2017,calcino2025a} and could explain at least parts of the flocculent spirals in AB~Aur \citep{calcino2025b}. The gravitational instability (GI) is also theorized to offer an explanation \citep{speedie2024}. However, the GI might be short-lived if irradiation from the central star is taken into account \citep{rowther2024}, so that it is natural to assume that both effects operate at the same time, with the infall supplying fresh material to keep the GI active \citep{longarini2025}. Well-defined spiral arms have also been shown to be the result of a stellar flyby \citep{quillen2005} or be caused by planets \citep{baruteau2014,muley2024} or gap-edge illumination \citep{muley2026}.

This raises the question of what distinguishes the signatures caused by different mechanisms and how to disentangle them. Furthermore, spiral patterns found in the scattered light with VLT/SPHERE do not always match with the spirals found in CO line observations with ALMA. For example, MWC~758 shows a one-armed spiral in the $^{12}$CO line emission \citep{winter2025}, while scattered light reveals two spiral arms \citep{benisty2015,ren2023}. In the millimeter continuum, the most prominent features are vortices, though a weak one-armed spiral continues to be visible \citep{dong2018}. For other disks, there might even be no spiral substructure in the continuum emission even if other types of observations hint at them, for example in SU Aur \citep{ginski2021}. Moreover, there can be discrepancies between the isotopologues of CO (e.g., \citealt{huang2025}), as these observations trace different disk layers. Infall could also be responsible for the formation of spiral structures even without direct evidence. For example, the MWC~758 system has no confirmed streamers. To explain its spiral structures, two planet candidates are proposed \citep{wagner2019,wagner2023} instead, but it is unclear whether these planets are connected to the spirals \citep{boccaletti2021}. A flyby is also an unlikely explanation in the case of MWC~758 \citep{grady2013,reggiani2018}, suggesting the possibility that infall may contribute to the formation of the spirals despite the absence of streamers.

In this work, we investigate the spiral structures that arise in select simulation setups of late infall by HD25, under which conditions they occur, how long they are present, and how they affect different layers of the disk. This work acts not only as a comparison to previous, similar works using SPH simulations \citep{calcino2025a}, but also as a more detailed characterization of the spiral structures that can be caused by late infall in the context of the vertical dependence of disk properties and substructures hinted at by the aforementioned observations. Allowing a comparison to observed systems also enables insights into the role the theorized environmental interactions play in forming them and which conditions provide a good match. A broader understanding of how and where disk substructures are shaped provides more clues on its impact on planet formation, which predominantly occurs close to the midplane, where large grains settle \citep{yl2007}, if the streaming instability \citep{yg2005} is the mechanism invoked for planetesimal formation (for a review, see \citealt{lesur2023}).

\section{Methods}\label{sec:methods}
We performed 3D hydrodynamical simulations on a spherical grid with logarithmic spacing in the radial direction, and uniform spacing in the polar and azimuthal direction. The computations were performed using the \texttt{FARGO3D} code \citep{benitez-llambay2016}, capable of GPU multiprocessing while utilizing the advection algorithm by \citep{masset2000}. We followed our approach from \citet[HD25]{hd25}, and refer interested readers to that article for a detailed description. In the following, we briefly summarize the main aspects of the simulations.

\subsection{Types of accretion models}
In HD25, we simulated the late accretion of gas onto a Class II disk, which is included as part of the initial condition, in two different scenarios. The first model is the capture of a cloudlet of gas, initialized in compact form on a hyperbolic orbit. We did not confine the cloudlet in any way. As a result, the cloudlet expands considerably before it encounters the disk, engulfing it in gas rather than creating a gas trail directly onto it. The goal of this more simplistic cloudlet capture setup is to model the encounter of the disk with an interstellar medium (ISM) overdensity in a controlled manner, and not to model the physical conditions in realistic detail. Initializing the cloudlet with a compact size has the advantage of keeping the inaccuracy incurred by treating it as a point source when calculating the initial orbital velocity low.

In the second model, we do not include any initial overdensities. Instead, we model a compressively turbulent background ISM with a systemic velocity with respect to the disk. This scenario represents a more realistic case of turbulent Bondi-Hoyle-Lyttleton (BHL) accretion onto the disk, where overdensities and related streamers emerge naturally as a result of the turbulence. As the disk moves through the ISM, an accretion tail emerges, and turbulent asymmetries result in net angular momentum relative to the star, creating streamers (see HD25 for details). This scenario is more realistic, but offers less control over the accretion process, which is mostly regulated by the infall rate, velocity dispersion, and turbulent power spectrum. We do not drive the turbulence; it is only included as an initial condition.

\subsection{Numerical setup}
We set the inner edge of the grid to $r_\mathrm{min}=\SI{5}{\astronomicalunit}$. For the cloudlet capture simulations, we employed $r_\mathrm{max}=\SI{5000}{\astronomicalunit}$, whereas we set $r_\mathrm{max}=\SI{50000}{\astronomicalunit}$ for all simulations with a turbulent ISM as initial condition. In the polar direction, the grid extended from $\theta_\mathrm{min}=\SI{10}{\degree}$ to $\theta_\mathrm{max}=\SI{170}{\degree}$. In the radial and polar direction, we used an outflow boundary condition; the azimuthal boundary was periodic. For the simulations of cloudlet capture, we chose $N_r=366$ cells in radial, $N_\theta=146$ cells in polar, and $N_\phi=330$ cells in azimuthal direction, whereas we used $N_r=486$ cells for the turbulent medium simulations. This results in a number of cells per scale height of $H/\Delta r=H/(r\Delta\theta)=H/(r\Delta\phi)\approx 2$ at $r=\SI{5.2}{\astronomicalunit}$ (${\approx}4.2$ at \SI{100}{\astronomicalunit}). Here, $H$ is the gas pressure scale height and $\Delta r$, $\Delta\theta$ and $\Delta\phi$ describe the resolution in the respective direction.

For the initial condition, we included a protoplanetary disk with a symmetric surface density profile of $\Sigma(r)=\Sigma_0\left(r/R_0\right)^{-a}$, with $R_0=\SI{5.2}{\astronomicalunit}$ and $a=1.5$. We applied an exponential cutoff at \SI{100}{\astronomicalunit}. For each accretion model, we considered two scenarios. First, a high disk mass, $\Sigma_0=\SI{310.547}{\gram\per\centi\meter\squared}$ resulting in $M_d=\SI{0.05}{\solarmass}$, and second, a low disk mass, $\Sigma_0=\SI{31.0547}{\gram\per\centi\meter\squared}$, resulting in $M_d=\SI{5e-3}{\solarmass}$. The simulations were performed using an isothermal equation of state and a solar-mass star, with an aspect ratio of $h(r)=h_0(r/R_0)^f$ with $h_0=0.03799$ and $f=0.25$. We included an $\alpha$-viscosity of \num{e-3} \citep{ss1973}, but we note that the numerical diffusion introduced by the grid also contributes to this value, and that it could be significant at our employed resolution. The floor density of the simulation was \SI{e-22}{\gram\per\centi\meter\cubed}.

\subsection{Accretion model initial conditions}
In the cloudlet scenario, we initialized the cloudlet with $v_\infty=\SI{.5}{\kilo\meter\per\second}$, $d_0=\SI{3000}{\astronomicalunit}$, $R_\mathrm{cloud}=\SI{500}{\astronomicalunit}$, and $M_\mathrm{cloud}=\SI{5e-3}{\solarmass}$. This results in an initial density of the cloudlet of $\rho_\mathrm{cloud}=\SI{5.67e-18}{\gram\per\centi\meter\cubed}$. The cloud initially expands due to the lack of pressure support, leading to a mean density of ${\sim}\SI{e-19}{\gram\per\centi\meter\cubed}$ at the time of the encounter with the disk, which is comparable to filamentary structures in the Taurus star forming region \citep{pineda2010,palmeirim2013}. The total simulated time of the cloudlet simulations is \SI{42.7}{\kilo\year}. We set the impact parameter to $b=0.5 b_\mathrm{crit}=0.5GM_\odot/v_\infty^2$. Different from the case considered in \citet[HD25]{hd25}, the orbit of the cloudlet is not inclined with respect to the plane of the disk, leading to in-plane capture. We chose this configuration because we suspect it to be most favorable for purely infall-driven spiral formation. First, the interaction between infalling material and gas is concentrated at the disk edges, so that the perturbation on the disk surface has the highest degree of asymmetry, promoting spiral formation. Second, the contamination of emission from the disk by the arising streamers and the accretion tail is minimal for this orientation. Third, the effect on the disk dynamics, for example, potential warping or tilting, which can also cause spiral patterns \citep{winter2025}, is also minimal for this configuration, allow us to separate the impact of the different formation pathways.

For the simulations with a turbulent medium, we considered a systemic infall rate of $\dot{M}_\mathrm{sys}=\SI{e-8}{\solarmass\per\year}$ and a systemic velocity of $v_\mathrm{sys}=\SI{0.5}{\kilo\meter\per\second}$, resulting in $\rho_\mathrm{ISM}=\SI{3e-21}{\gram\per\centi\meter\cubed}$. Unlike HD25, we chose the direction of the systemic velocity such that it leads to in-plane streamers, that is, the systemic velocity vector and angular momentum vector of the disk are perpendicular, $\hat{\vec v}_\mathrm{sys}=(-1,0,0)$, for the same reason as for the cloudlet capture case. We included both small and large scales in the turbulence power spectrum, with $k_\mathrm{min}=2\pi\times\SI{e-5}{\per\astronomicalunit}$ and $k_\mathrm{max}=2\pi/50~\si{\per\astronomicalunit}$. The turbulent velocity dispersion is $\sigma_\mathrm{turb}=\SI{0.5}{\kilo\meter\per\second}$. The simulated time of the turbulent medium simulations is \SI{30}{\kilo\year}.

\subsection{Synthetic observations}
Using the Monte Carlo radiative transfer code \texttt{RADMC3D} \citep{radmc3d}, we created synthetic observations of select simulation snapshots, as in HD25. We produced observations of the $^{12}$CO and $^{13}$CO J 2--1 transition line, assuming a constant $^{12}$CO abundance of $\num{e-4}$ and a $^{13}$CO abundance of $\num{e-4}/70$ \citep{wr1994}, using 250 points in frequency space covering a spectral range from \SI{-5}{\kilo\meter\per\second} to \SI{5}{\kilo\meter\per\second} to create first moment maps. Spatially, we used 400 pixels for each axis, covering a range from \SI{-100}{\astronomicalunit} to \SI{100}{\astronomicalunit}. As in HD25, we did not employ any prescriptions for the freeze-out or photo-dissociation of the CO isotopologues and re-computed the temperature using \texttt{RADMC3D} instead of using the isothermal temperature distribution from the simulations. We assumed $T_\mathrm{gas}=T_\mathrm{dust}$.

The main focus of this work is on synthetic scattered-light observations. Here, the required dust opacities were calculated using the \texttt{optool} code \citep{dominik2021}, where we assumed the DIANA standard dust model composition \citep{woitke2016}. We assumed a porosity of 25\% and mix 15 different grain sizes between \SI{1}{\micro\meter} and \SI{3}{\micro\meter}, sampling an $n(a)\propto\ a^{-2.5}$ power-law size distribution. We set the dust-to-gas volume density ratio to 1\%, which holds true under the assumption that the grain size distribution on the disk surface, perturbed by the infall, is predominantly given by small grains, because large grains are removed from the surface due to settling, and new small grains are constantly replenished from the ISM. This assumption would not hold for the disk midplane, where the large grains are present and contain a majority of the mass, but these disk regions are not probed by the synthetic scattered light observations. For better numerical treatment of the scattering peak, we applied ``chopping'' of \SI{2}{\degree} \citep{dominik2021}, which is shown to be a reasonable assumption in similar conditions of protoplanetary disks \citep[see][Appendix~A]{kimmig2025}. To enable the study of the scattered light emission as a function of time, analyzing many simulation snapshots, we utilize that simple scattering approximation included in \texttt{RADMC3D}, which enable direct integration of the radiative transfer problem by taking advantage of the specific problem geometry. It assumes that all scattered light is starlight, that thermal emission can be neglected, and that multiple scattering does not occur. For all direct depictions of scattered light emission shown in subsequent sections, we compared this approximation to a full Monte-Carlo model and found no qualitative difference.

\subsection{Determination of residual motions}
We characterize the spiral structures found using the synthetic CO line emissions by considering the residual motions, that is, the deviation of the determined moment~1 velocities from a Keplerian disk model. While we could have, in principle, full knowledge of the emission from the unperturbed disk by considering the initial condition, key parameters that are relevant for an observation determination of an accurate Keplerian model are altered due to the accretion that occurs over the course of the simulation. For example, the apparent emission height, inclination, and position angle are affected. To maintain comparability to observations, we therefore discard any information obtained from the knowledge of the initial conditions, and fit a Keplerian model purely based on the synthetic observation at the considered time.

The fit is performed using the Monte Carlo disk fitting code \texttt{discminer} \citep{izquierdo2021}. In order to isolate the disk emission from cloud contamination as much as possible, we restrict the disk model to a maximal size of \SI{100}{\astronomicalunit}, and exclude emission from the inner \SI{10}{\astronomicalunit}. In the interest of computational cost, we downsample the synthetic image by six pixels, and we only consider emission from within \SI{150}{\astronomicalunit}. Therefore, only emission at a distance $\SI{10}{\astronomicalunit}<r<\SI{150}{\astronomicalunit}$ was considered. We perform the fit using 250 walkers that take 10000 steps while fixing the stellar mass to \SI{1}{\solarmass}, enforcing a centered disk, and fixing the systemic velocity at zero. 

\section{Results}
In this work, we investigate the formation of spiral structures, mainly those on the disk surface which are probed by scattered light observations, in simulations with setups similar to those in HD25. In particular, we consider a cloudlet capturing onto a disk with a mass of \SI{5e-2}{\solarmass} in-plane (simulation~1) and BHL accretion with an accretion rate of \SI{e-8}{\solarmass\per\year} onto a disk of the same mass (simulation~2). We also consider these scenarios for a disk with a mass of \SI{5e-3}{\solarmass} (simulations 3 and 4). We put the main emphasis on simulation~1, and explore how a different infall model, or infall onto a lighter disk, changes the picture in subsequent sections.

\subsection{Cloudlet capture}
\begin{figure*}[htp]
    \centering\includegraphics[width=.8\linewidth]{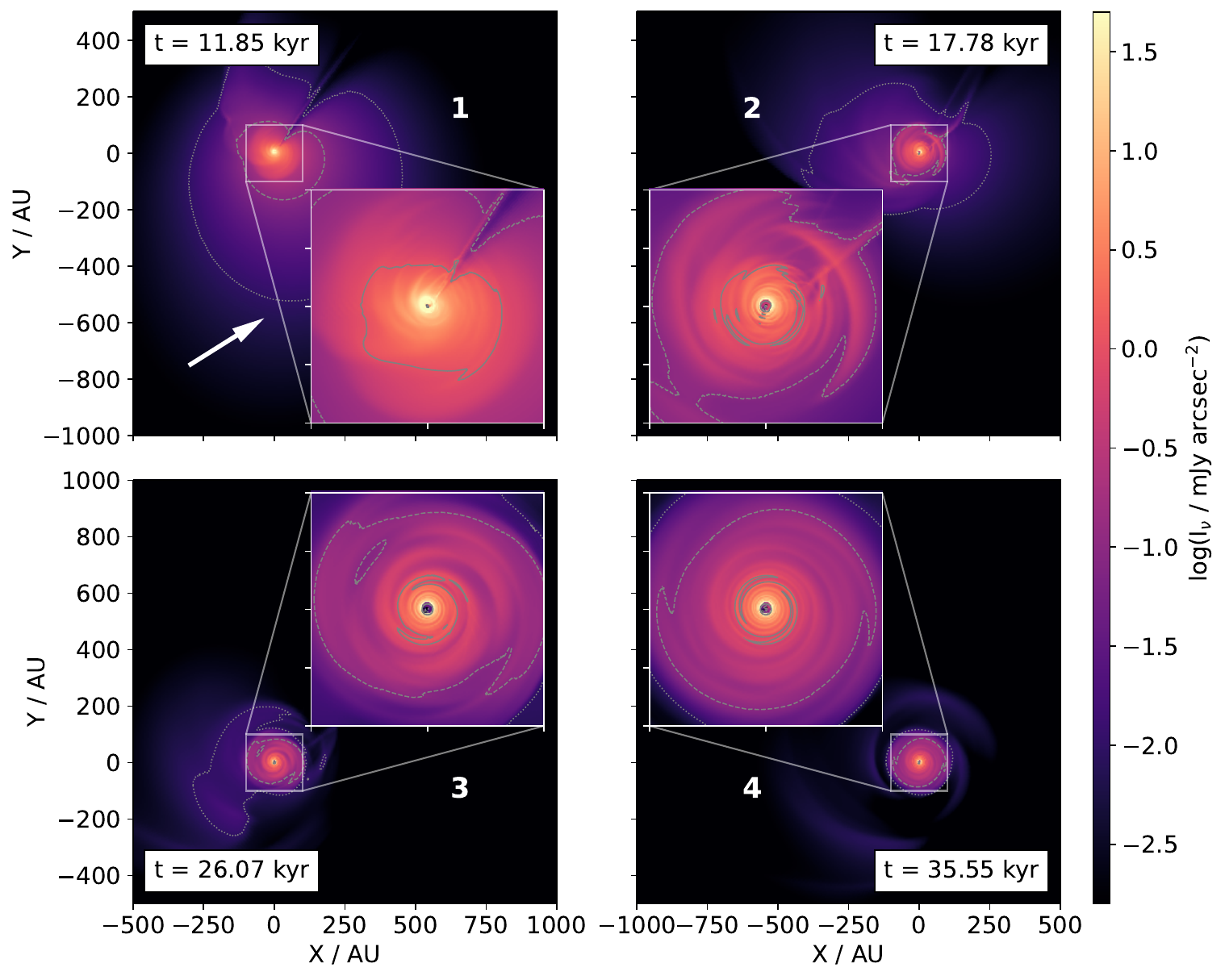}
    \caption{Polarized scattered light intensity in four snapshots of simulation~1. The solid gray line denotes the \SI{1}{\milli\jansky\per\arcsecond\squared} contour, the gray dashed line is the \SI{0.1}{\milli\jansky\per\arcsecond\squared} contour, and the gray dotted line the \SI{0.01}{\milli\jansky\per\arcsecond\squared} contour. The panel numbering corresponds to the phases of the cloudlet encounter. The white arrow direction denotes the original orbital velocity of the cloudlet.}
    \label{fig:cldl_spirals_rphi}
\end{figure*}%
The capture of a cloudlet of gas leads to various spiral structures at different times in the simulation. These structures, either spirals that have a low number of arms or that are very flocculent, can be characterized depending on the environmental conditions at that time, and whether they occur in the inner or in the outer disk regions. We show synthetic polarized scattered light observations for four representative snapshots in Fig. \ref{fig:cldl_spirals_rphi}. The camera angle is chosen such that the disk appears fully face-on. We identify four different phases in our simulation. The first phase is characterized by the initial, main encounter of the cloudlet with the disk, before overshooting gas falls back and streamers form. Here, we find little structure in the outer disk, whereas the inner disk shows several flocculent spiral arms. These flocculent spiral arms persist once the fallback streamer arises, which we call the second phase, but they are considerably different in shape and have fainter emission. At this stage, we also find a strong two-armed spiral, which we refer to as an $m=2$ mode, emerging in the outer disk for the first time in the simulation.

We find that another $m=2$ spiral structure arises after the fallback streamer has largely dissipated, which we classify as the third phase. This structure is more clearly visible than the one during phase 2. It stretches from inner to the outer disk and is not accompanied by flocculent structures. At this time, the bulk of the mass has already been accreted; this structure is therefore solely caused by the accretion of remnant gas leftover from the cloudlet encounter. Here, the spiral structure is also directly reflected in the contour lines of the emission (see Fig.~\ref{fig:cldl_spirals_rphi}, bottom left panel). Qualitatively, the spiral persists for a long time (${\sim}\SI{20}{\kilo\year}$), but it stops extending to the outer disk after ${\sim}\SI{10}{\kilo\year}$, leaving behind a structure in the inner disk that appears to be more wound-up (see phase 4 in Fig. \ref{fig:cldl_spirals_rphi}).

\begin{figure*}[htp]
    \centering\includegraphics[width=.9\linewidth]{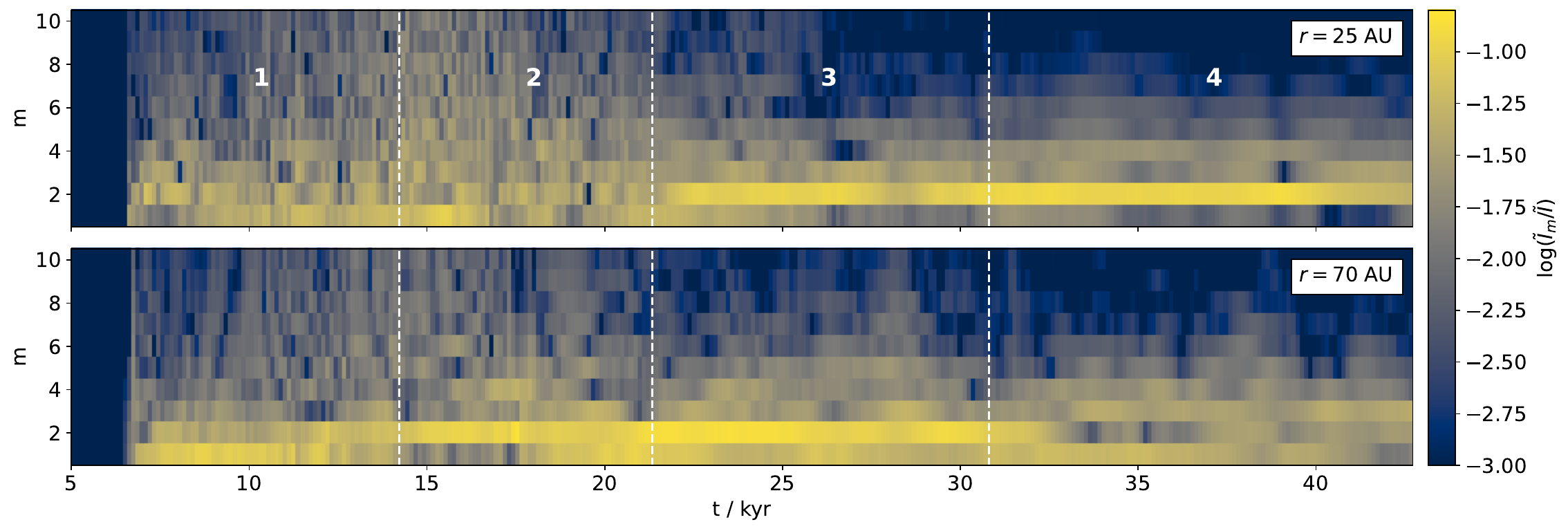}
    \caption{Azimuthal Fourier decomposition of the polarized scattered light intensity, normed by the total intensity. $m=\tilde{\phi}$ is the Fourier transform of the azimuthal coordinate, and the two panels show the transformation for two radii. The white dashed lines separate the phases of the encounter, labeled by the white numbers.}
    \label{fig:cldl_spectrum}
\end{figure*}%
To quantify the strength of the individual spiral modes, we perform an analysis of the Fourier-transformed azimuthal intensity distribution, shown in Fig. \ref{fig:cldl_spectrum}, where $m=\tilde{\phi}$ is the Fourier transform of the azimuthal coordinate. The transformation is performed at two distinct radii that we find to be representative of structure in the inner ($r=\SI{25}{\astronomicalunit}$) and outer ($r=\SI{70}{\astronomicalunit}$) disk. The classification into the four phases is based on the excitation of the Fourier modes, as shown in Fig.~\ref{fig:cldl_spectrum}. This figure confirms the previous findings, where the amplitude of the $m=2$ mode is particularly strong in the inner disk starting at phase 3, and the first $m=2$ mode with strong amplitude is found in the outer disk during phase 2. The presence of flocculent spiral structure can be seen in phases 1 and 2, where also higher-order modes are excited. The Fourier analysis suggests that the $m=1$ mode is also excited in the outer disk during phase 1. The excitation of this mode is not related to the presence of an $m=1$ spiral, but rather caused by the contamination of the scattered light image from the accretion tail caused by the cloudlet interacting with the disk.

\subsubsection{Pattern speed}
\begin{figure*}[htp]
    \centering\includegraphics[width=.9\linewidth]{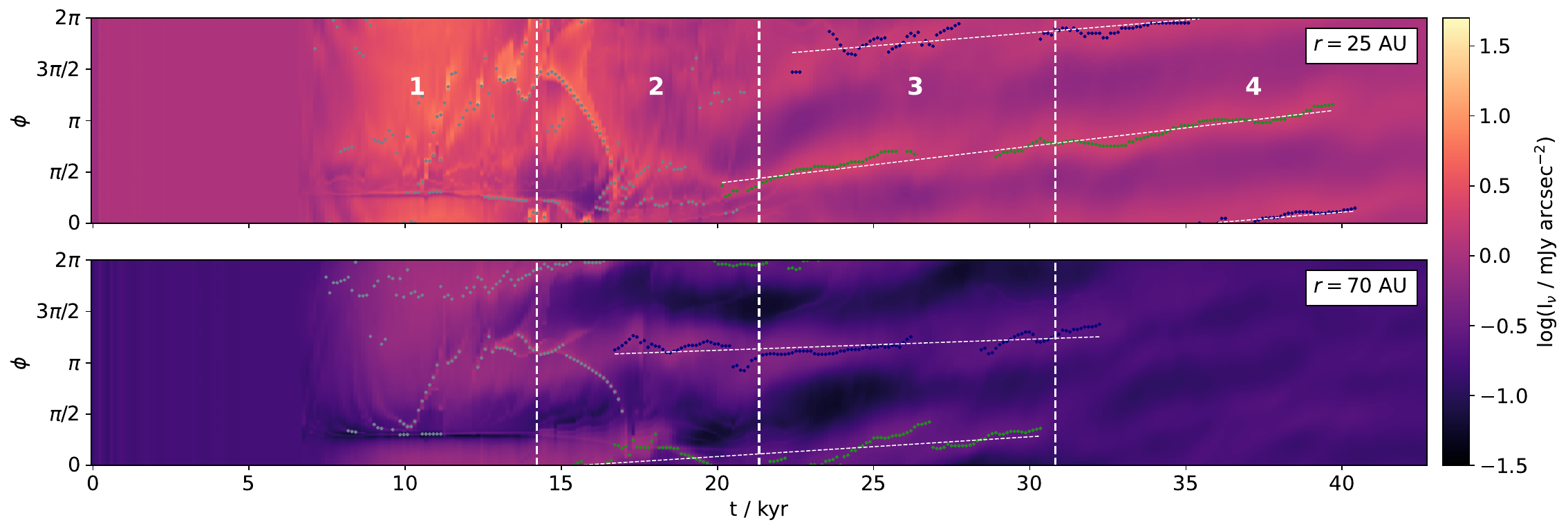}
    \caption{Polarized scattered light intensity as a function of azimuth and time for two different radii. The white dashed lines and numbers denote the different encounter phases, as in Fig.~\ref{fig:cldl_spectrum}. The colored dots show the azimuthal peaks, colored in gray if not used for further consideration, otherwise by membership to one of the spiral arms. The solid white lines show the linear fit used to determine the pattern speed.}
    \label{fig:cldl_spirals_tphi}
\end{figure*}%
In an effort to characterize the $m=2$ spirals found across the majority of the disk in phases 3 and 4, we calculated the pattern speed, an observable spiral property that is particularly useful if the spirals are caused by a perturber, for example, a planet or another star. We utilize the peak finding algorithm from the \texttt{scipy} Python package to find the location of the spirals in azimuth at the same radii as for the modal analysis above, that is, $r\in\{25,70\}~\si{\astronomicalunit}$. In other words, we find all peaks in two rings at each point in time. The algorithm is applied to the logarithmic intensity values, which are divided by the median initial intensity at each radius. We require that a peak is brighter than the initial median intensity and has a topographic prominence\footnote{The topographic prominence is defined as the lowest distance required to descent from the peak before a higher peak can be reached.} of 0.3. The results are marked in Fig. \ref{fig:cldl_spirals_tphi}. Afterwards, we use the clustering algorithm \texttt{DBSCAN} from the Python package \texttt{sklean} to separate the two spiral arms. We only consider peaks that we determine visually to be part of the $m=2$ spiral (marked in green and blue in Fig.~\ref{fig:cldl_spirals_tphi}), and discard all others (marked in gray in Fig.~\ref{fig:cldl_spirals_tphi}).

The pattern speed was then determined through a linear fit of the azimuthal peaks over time, as shown in Fig.~\ref{fig:cldl_spirals_tphi}. We find that the spirals are close to stationary. For the inner disk, the arm marked in blue has a pattern speed of $\dot{\phi}=\SI{0.08}{\per\kilo\year}$, corresponding to a corotation radius {$r_\mathrm{corot}=\SI{1840}{\astronomicalunit}$} and the green one has $\dot{\phi}=\SI{0.11}{\per\kilo\year}$ ({$r_\mathrm{corot}=\SI{1460}{\astronomicalunit}$}). The pattern speed is especially slow in the outer disk. Here, the blue spiral arm moves with $\dot{\phi}=\SI{0.05}{\per\kilo\year}$ ($r_\mathrm{corot}=\SI{3150}{\astronomicalunit}$) and the green one with $\dot{\phi}=\SI{0.07}{\per\kilo\year}$ ($r_\mathrm{corot}=\SI{2040}{\astronomicalunit}$). This result acts as a defining difference to spirals created by other mechanisms, where either a Keplerian pattern speed or one corresponding to a specific perturber are expected, so that the inferred {$r_\mathrm{corot}$} would be smaller.

Measurements determining spiral pattern speeds have been conducted observationally. For the spiral in the disk around SAO~206462, for example, a pattern speed corresponding to {$r_\mathrm{corot}=\SI{86}{\astronomicalunit}$} was measured \citep{xie2021}. Additionally, in V1247~Ori, the pattern speed corresponds to $r_\mathrm{corot}=\SI{118}{\astronomicalunit}$. These observed values are much smaller than what we find in our simulations. Thus, this can serve as a way to distinguish spirals formed purely through infall from those formed through other mechanisms. We confirmed the low pattern speed of the spirals in our simulations not to be an effect of the output frequency.

\subsubsection{Gas kinematics}\label{sec:cldl_co}
\begin{figure}[htp]
    \centering\includegraphics[width=\linewidth]{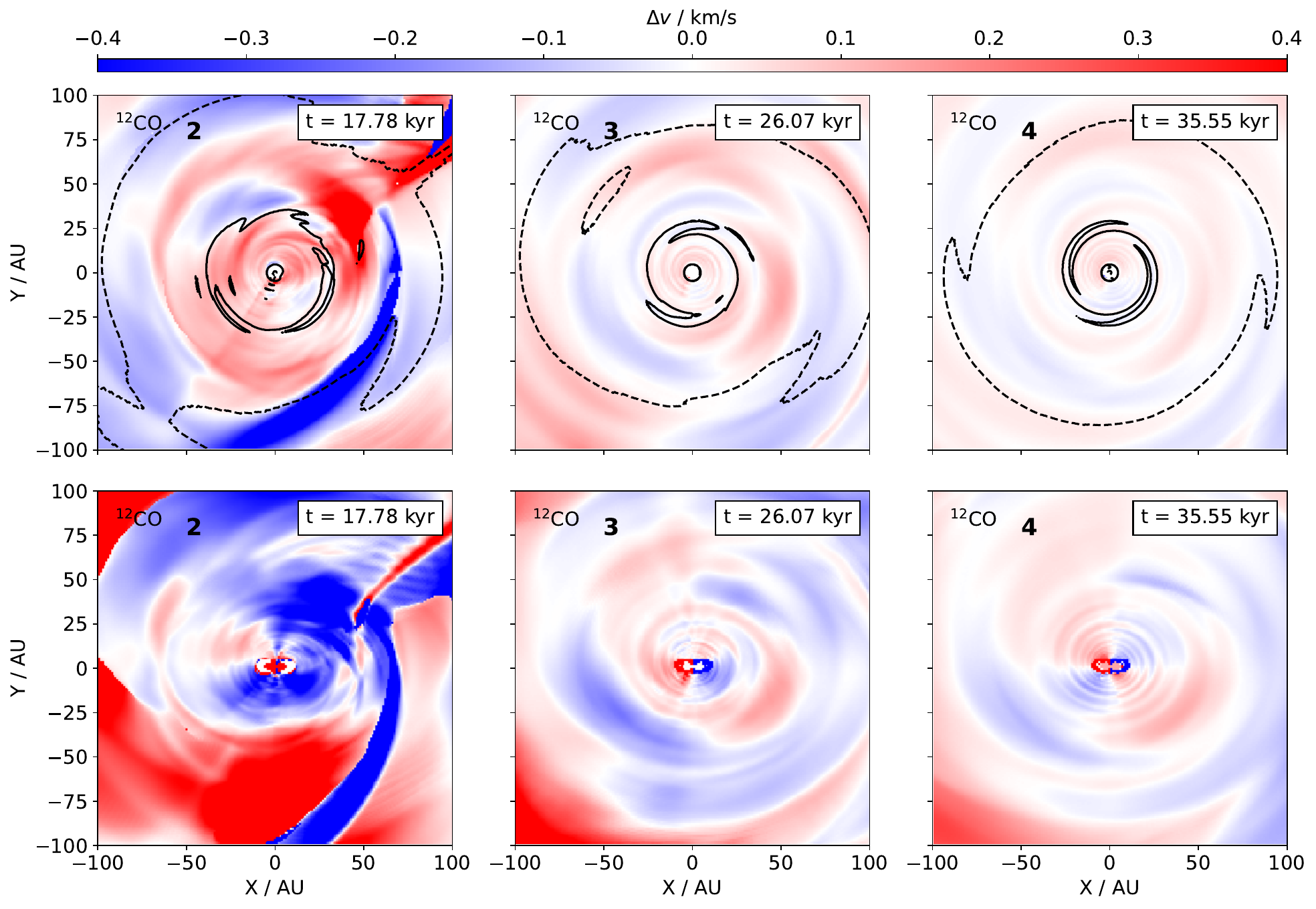}
    \caption{Moment 1 residuals from Keplerian motion of the $^{12}$CO line emission. The top row shows three snapshots, taken at the same time as in Fig.~\ref{fig:cldl_spirals_rphi}, where the camera is oriented face-on. The bottom row instead shows images where the camera has an inclination of \SI{30}{\degree}. The black lines show polarized scattered light contours, as in Fig.~\ref{fig:cldl_spirals_rphi}. The panel numbers correspond to the evolutionary phase.}
    \label{fig:cldl_co}
\end{figure}%
\begin{figure}[htp]
    \centering\includegraphics[width=\linewidth]{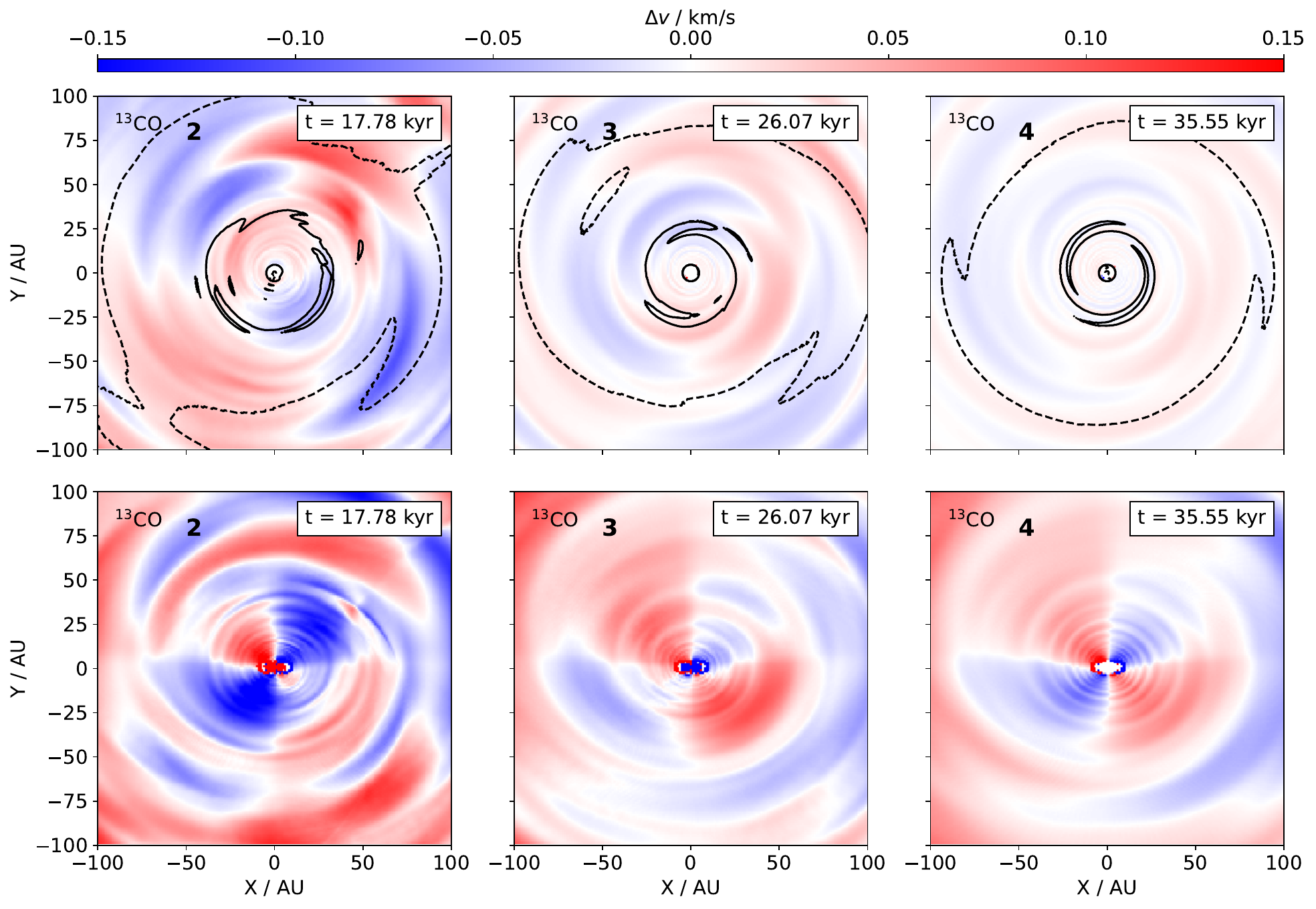}
    \caption{Same as Fig.~\ref{fig:cldl_13co}, but for $^{13}$CO. We note that the value range of the color bar is reduced.}
    \label{fig:cldl_13co}
\end{figure}%
Substructures found in polarized scattered light trace the density structure of the disk surface layer. In order to obtain information about the kinematical structure and deeper disk layers, we investigated the gas kinematics by creating synthetic observations of $^{12}$CO and $^{13}$CO J 2--1 line emission, and considering the first moment of that emission. Due to the fact that the disk is surrounded by gas from the environment, and that we neglect possible photodissociation of CO, the emission of $^{12}$CO is very optically thick, and the $\tau=2/3$ surface, where $\tau$ is the optical depth, is high above the midplane (see Fig. \ref{fig:cldl_depth} below). For the $^{13}$CO emission on the other hand, the $\tau=2/3$ surface lies deeper inside the disk, where perturbations created by the infall are smaller. We investigated deviations from a purely Keplerian rotation, which would correspond to a fully unperturbed, isolated disk, by computing the residuals of the first moment map using the \texttt{discminer} code. The usage of its Monte-Carlo fitting routines allows for the determination of the apparent disk inclination and position, which is different from the expected orientation given by the camera angle, as possible asymmetries and changes in disk tilt can be introduced during the accretion. The determined residuals are shown in Fig.~\ref{fig:cldl_co} for $^{12}$CO emission, and Fig.~\ref{fig:cldl_13co} for $^{13}$CO emission. The \texttt{discminer} fit parameters can be found in Tables \ref{tab:discminer_1_2}, \ref{tab:discminer_1_3}, and \ref{tab:discminer_1_4}.

The optically thick $^{12}$CO emission, tracing the upper layers of the disk, shows spiral structures in its residuals. They are most apparent during phases 3 and 4, at the time when the $m=2$ spiral is prominent in scattered light. We find that the physical location of the spirals in the disk does not coincide with their location in scattered light. However, the spiral shape matches between the $^{12}$CO and scattered light emission. The spirals are most prominent when the disk is viewed face-on, suggesting that they are caused by vertical motion induced by the cloudlet capture. In contrast, the residuals we find when the disk is viewed at a \SI{30}{\degree} angle reveal a more large-scale structure in the outer disk, which we interpret to be a one-armed spiral for phase 2, and a two-armed one for phases 3 and 4.

We find that the emergence of substructures in the CO emission depends strongly on the viewed isotopologue. For $^{13}$CO, residuals of considerable amplitude are only found during phase 2, where an active streamer is falling onto the disk. The distribution of substructures matches that of the $^{12}$CO emission, but at a lower amplitude. For the inclined view, a spiral is visible in phase 2, but cannot be found in the later phases. We note that the residuals shown in these later phases are the result of a model limitation in the \texttt{discminer} code, which is related to the characterization of the emission surface and errors introduced by including environment emission in the fit of the Keplerian model.

\subsubsection{Perturbation depth}\label{sec:cldl_depth}
\begin{figure*}[htp]
    \centering\includegraphics[width=.9\linewidth]{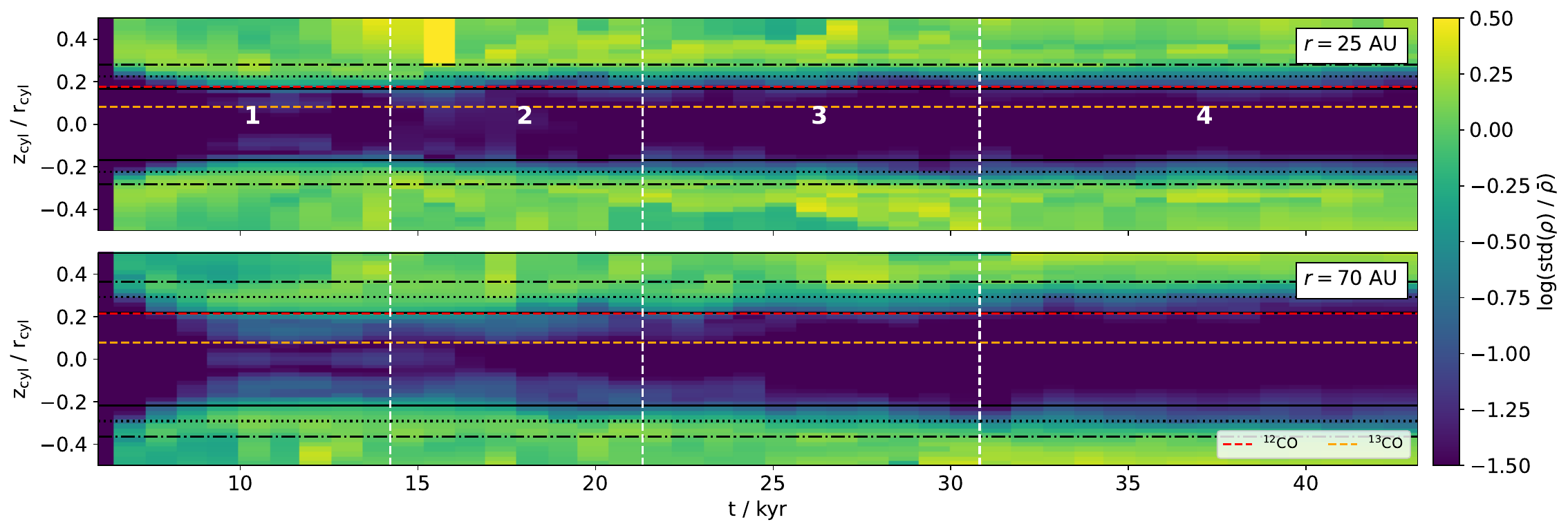}
    \caption{Coefficient of variation ($\sigma_\rho/\bar{\rho}$) in azimuth as a function of time and height above the disk midplane. The white dashed lines denote the different phases of the encounter. The black lines represent three (solid), four (dotted), and five (dash-dotted) pressure scale heights. The coefficient is calculated for $r=\SI{25}{\astronomicalunit}$ and $r=\SI{70}{\astronomicalunit}$, shown in the top and bottom panel, respectively. The red, orange, and yellow dashed lines denote the median in time of the $\tau=2/3$ emission surface of $^{12}$CO and $^{13}$CO, respectively.}
    \label{fig:cldl_depth}
\end{figure*}%
The amplitude of the visible structures in the residual motions strongly depends on the isotopologue. This indicates a weak overall influence of the infall on disk kinematics, meaning that strong perturbations only occur in the upper disk layers, where the density is low. This result is expected, as the total mass of the cloudlet is only 10\% of the disk mass. In Fig.~\ref{fig:cldl_depth}, we characterize the perturbation depth in more detail by calculating the azimuthal ``coefficient of variation'' (CV) of the density, $\sigma_\rho/\bar{\rho}$, over time as a function of height above the midplane. This coefficient is a measure for the strength of the density perturbations at a given height. For the calculation, we use the same two radii as for the calculation of the azimuthal Fourier decomposition. As the residual motions for the two different CO isotopologues suggest, perturbations in the density strongly decrease in layers below three pressure scale heights in the inner disk, and only reach substantial values in the outer disk during the initial cloudlet encounter (phase 1). In fact, CV values above $0.1$ are not reached below four pressure scale heights in the inner disk after phase 2, when the $m=2$ spiral arises. This means that the infall-induced density fluctuations below this height are smaller than 10\% of the mean density. The effect of a deeper perturbation on the emergence of spiral patterns seems to switch between the inner and outer disk. In the inner disk, the $m=2$ spiral is only visible in phases 3 and 4, were the perturbation depth is minimal. In contrast, in the outer disk, the $m=2$ spiral disappears for the minimal depth in phase 4, but is strong during the deeply perturbed phases 2 and 3.

\subsubsection{Effect on orbital elements}
\begin{figure*}[htp]
    \centering
    \begin{minipage}[t]{.65\linewidth}
        \vspace*{0pt}
        \includegraphics[width=\linewidth]{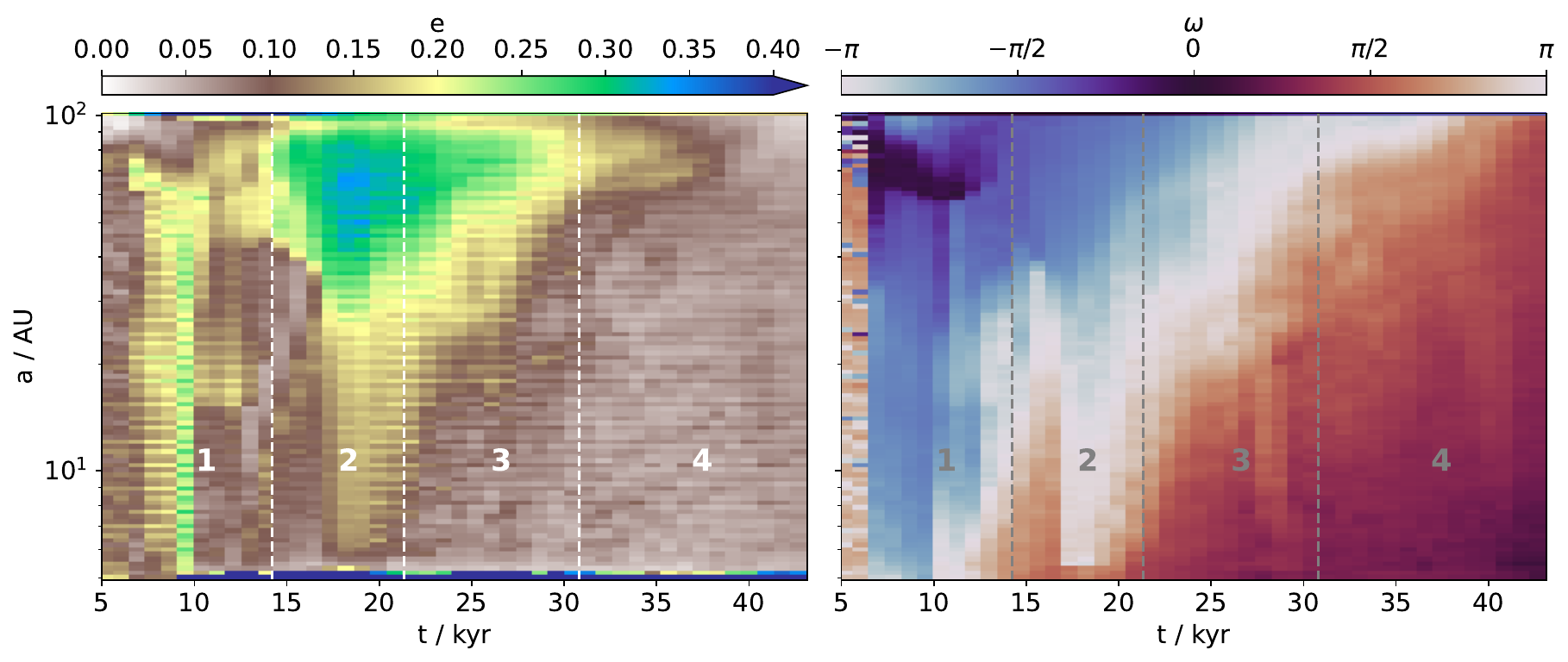}
    \end{minipage}%
    \begin{minipage}[t]{.25\linewidth}
        \vspace*{.5cm}
        \includegraphics[width=\linewidth]{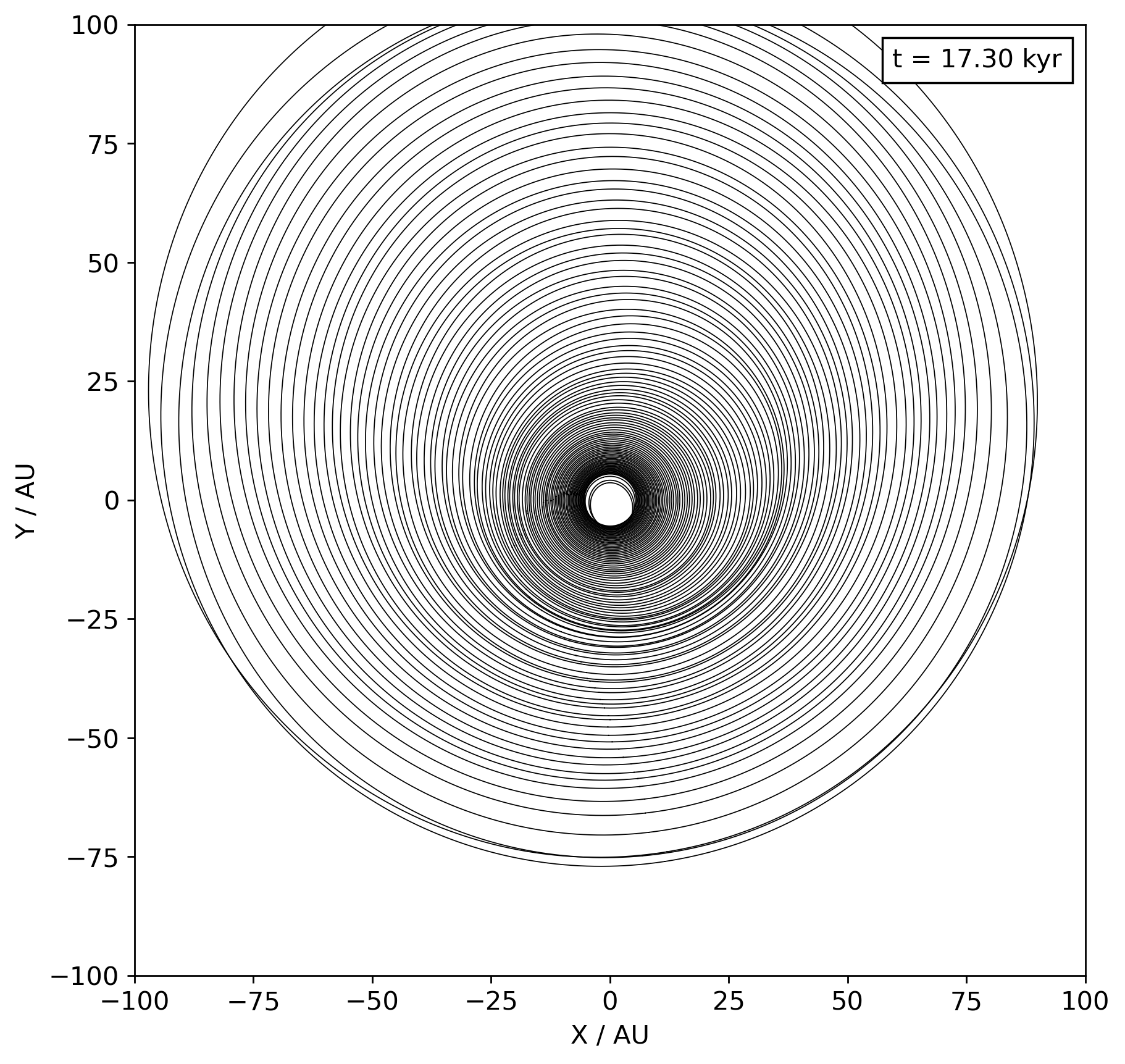}
    \end{minipage}%
    \caption{Volume-weighted averages of the orbital elements of the gas as a function of time and semi-major axis. The left panel shows the eccentricity, while the middle panel shows the argument of periapsis. The vertical dashed lines show the different phases of the encounter. The right panel shows the elliptical orbits for each semi-major axis bin, viewed face-on, using the $e$ and $\omega$ values from the other two panels.}
    \label{fig:cldl_orbit}
\end{figure*}%
In an effort to shed more light on the mechanism that leads to the formation of the spirals, we calculated the orbital properties of the gas in the directly influenced upper layers. We determined the eccentricity and argument of periastron as a function of semi-major axis, which was split into 100 bins. Values from cells whose semi-major axis values lie within the same bin were averaged, and values were weighted by cell volume. We chose not to use the density or mass of the cell as weight, as this would favor the unperturbed disk midplane. As we are mainly interested in surface-level effects, using the cell volume as weight is a choice that puts emphasis on the upper disk layers. To prevent cells that are unrelated to the disk itself from influencing the calculated average, we only consider cells that are within 4 pressure scale heights from the disk midplane.

The resulting values, as a function of time, are shown in Fig.~\ref{fig:cldl_orbit}. We find an excitation of eccentricity in the outer disk, up to $e\sim 0.35$, once the fallback streamer emerges. Similar to recent results by \citet{calcino2025a}, where twisted elliptical orbits acted as an explanation for the spirals arising in their model of cloudlet capture, we find a shift in the argument of periastron at the locations with excited eccentricity. However, unlike their findings, the resulting elliptical orbits do not align in a way that would promote the formation of low-modal spirals for the majority of the time. The most prominent, but brief, spiral structure that forms as a result of that mechanism is shown in the right panel of Fig.~\ref{fig:cldl_orbit}. It is an $m=1$ spiral in the inner disk, which is not related to any structures visible in scattered light. This indicates that the spirals found in our simulation are a direct consequence of the late infall accretion, and not a consequence of perturbed disk dynamics. We note that, in principle, the induced eccentricity subjects the upper disk layers to a parametric instability, which could drive vertical flow patterns and turbulence \citep{ogilvie2014,dewberry2025}.

\subsection{Accretion in a turbulent medium}\label{sec:bhlh}
\begin{figure*}[htp]
    \centering\includegraphics[width=.8\linewidth]{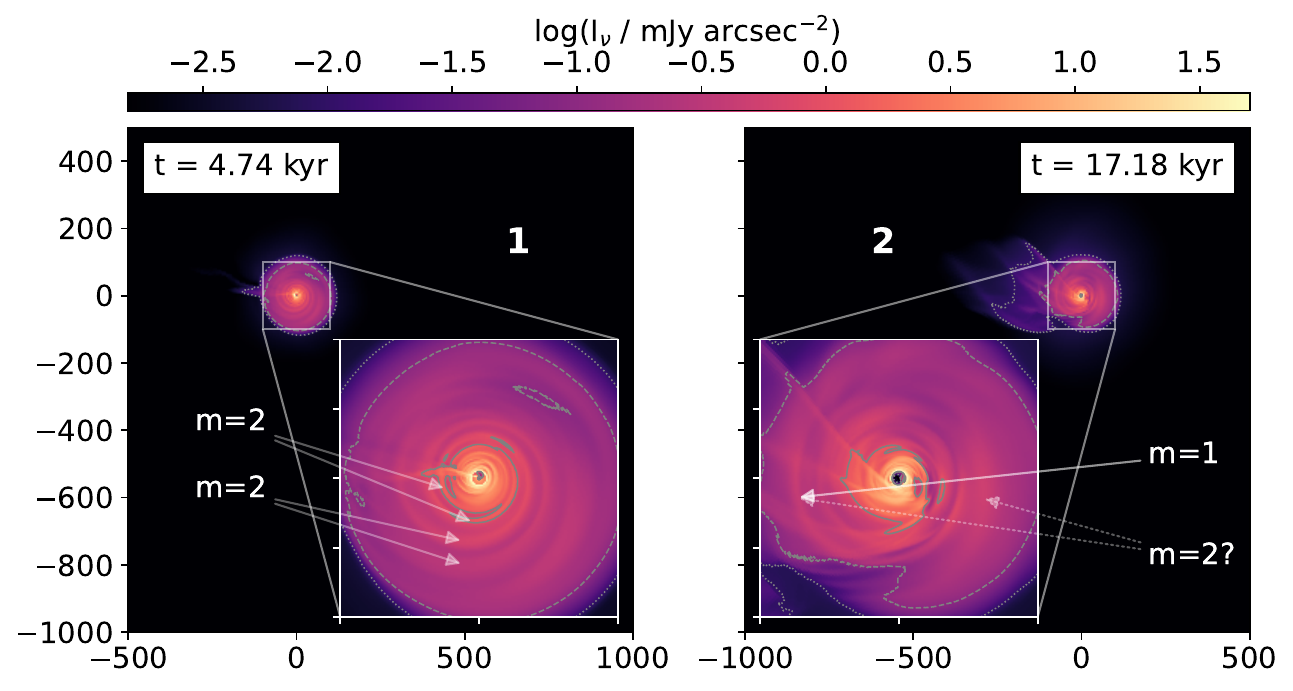}
    \caption{Same as Fig.~\ref{fig:cldl_spirals_rphi}, but for simulation 2. The white arrows denote the number of spiral arms $m$ determined by eye. For the right panel, the dotted arrow denotes an alternative interpretation where the $m=1$ spiral is an $m=2$ spiral, but one arm is much fainter.}
    \label{fig:bhlh_spirals_rphi}
\end{figure*}%
If late infall occurs in the form of BHL accretion, that is, gas is accreted from a turbulent environment, the emerging spiral structure is considerably different from that in the cloudlet capture scenario. Only two different kinds of structures arise, which we again use to divide the evolution into two phases, shown in Fig.~\ref{fig:bhlh_spirals_rphi} using two representative snapshots\footnote{We note that the physical times of the evolutionary phases in the BHL accretion case do not match those of Fig.~\ref{fig:cldl_spirals_rphi}.}. In phase 1, a faint Bondi-Hoyle accretion tail is present, but no separate streamer has formed yet. Here, an $m=2$ spiral can be seen, especially in the outer disk. The distinct spiral arms are marked in Fig.~\ref{fig:bhlh_spirals_rphi}. Once a streamer has formed, which we classify as phase 2, the inner disk $m=2$ spiral can no longer be seen due to the perturbations at the disk surface, and the scattered light from the streamer arms make it difficult to discern clear spiral arms by eye. A spiral structure with a low modal number remains in the outer disk, although the exact dominating mode is also hard to discern by eye: It could be a one-armed spiral, or a two-armed spiral where one arm is considerably dimmer than the other.

\begin{figure*}[htp]
    \centering\includegraphics[width=.9\linewidth]{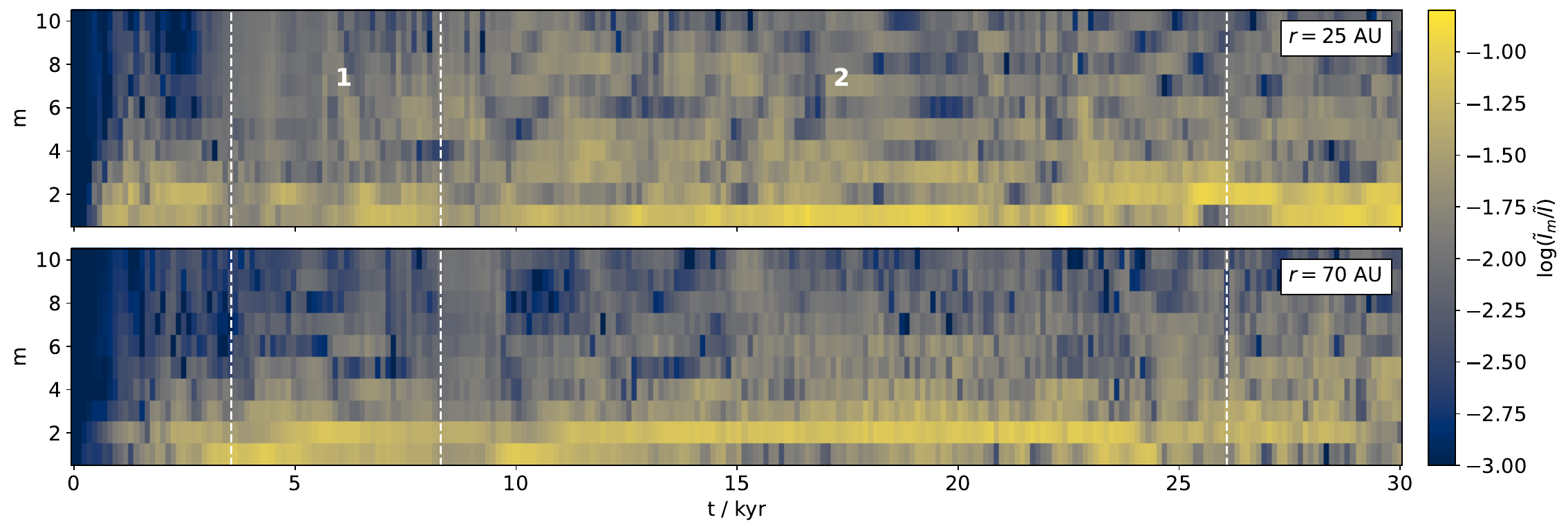}
    \caption{Same as Fig.~\ref{fig:cldl_spectrum}, but for simulation 2, where we model a BHL accretion scenario.}
    \label{fig:bhlh_spectrum}
\end{figure*}%
Analogously to the analysis of the cloudlet capture case, we analyzed the strength of the individual spiral modes in Fourier space, shown in Fig.~\ref{fig:bhlh_spectrum}. Here, we identify the two evolutionary phases mentioned above. The initial $m=2$ spiral in the outer disk during phase~1 is less strong in the inner disk, and frequently disappears. Both the inner and outer disk exhibit a strong $m=1$ mode, although we suspect this to be caused by the contamination of the scattered light image by the Bondi-Hoyle tail. During phase~2, when the arced streamer is present, the Fourier decomposition is dominated by the $m=1$ mode in the inner disk, and the $m=2$ mode in the outer disk. Since the streamer appears fainter further from the star, we conclude that the spiral structure present in the outer disk at that time is likely to be truly dominated by an $m=2$ mode, and not a result of contamination. However, similar to phase 1, we suspect contamination by the direct signature of the late infall to play a significant role for the inner disk. To examine this further, we explore whether this $m=1$ mode may indeed be a disk-inherent structure in the subsequent sections.

A fundamental difference of this accretion scenario compared to cloudlet capture is the continued excitation of higher order modes. This is caused by the sustained presence of a streamer, unlike the cloudlet capture scenario, where the streamer only persist for a short time, and eventually dissipates, leading to accretion of the gas remnant. As a result, the Fourier decomposition of the azimuthal intensity profile calculated for the BHL accretion case is more prone to contamination by the streamer.

\subsubsection{Gas kinematics}
\begin{figure}[htp]
    \centering\includegraphics[width=\linewidth]{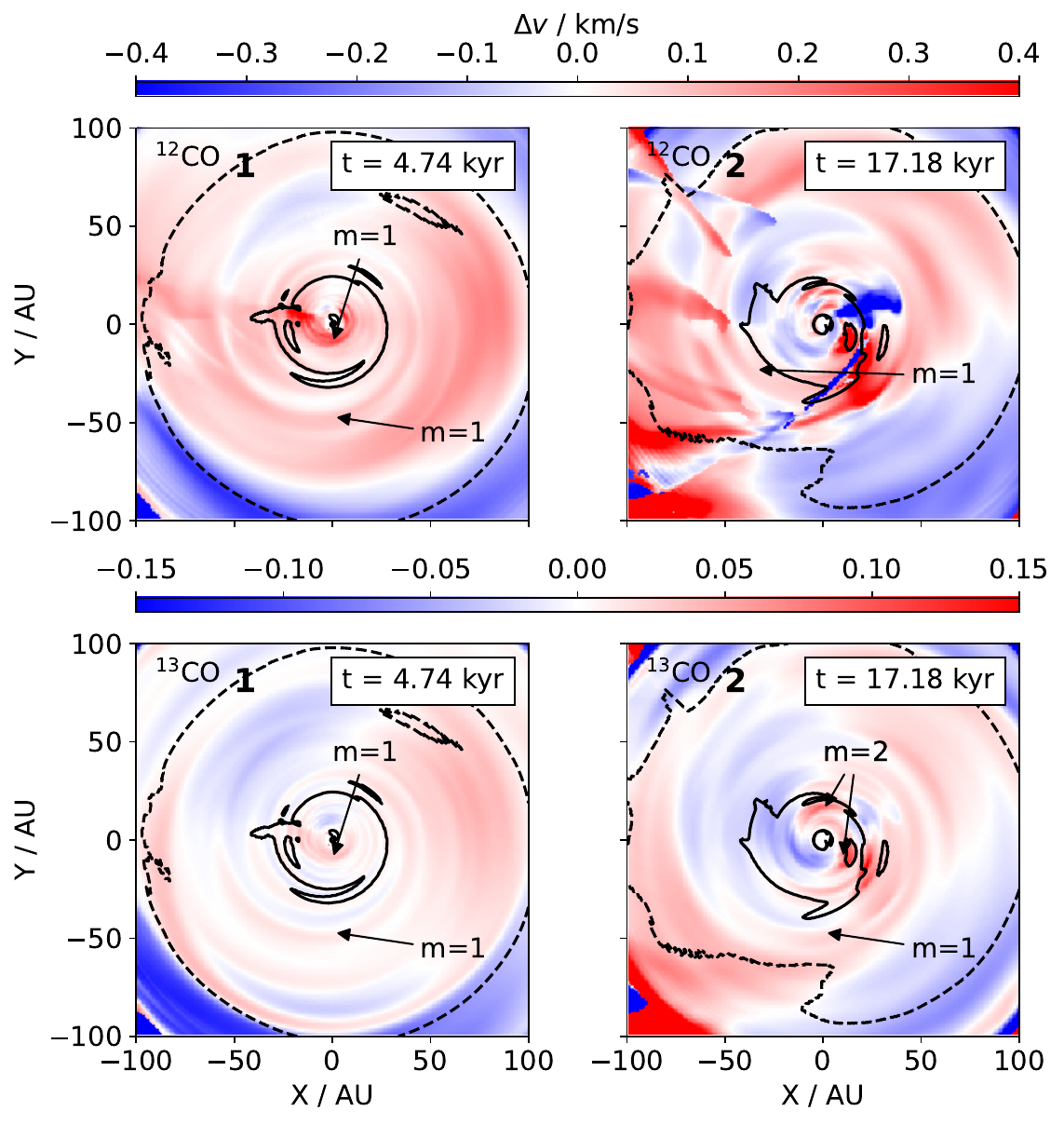}
    \caption{Moment 1 residuals from Keplerian motion of the $^{12}$CO and $^{13}$CO isotopologues for simulation 2. Top: Same as the top row of Fig.~\ref{fig:cldl_co}. Bottom: Same as the top row of Fig.~\ref{fig:cldl_13co}. The arrows denote the number of spiral arms $m$ determined by eye.}
    \label{fig:bhlh_co}
\end{figure}%
We further investigate how the gas kinematics are affected in this scenario. Analogously to Section \ref{sec:cldl_co}, we compute the residual motion found in the moment 1 map of the $^{12}$CO and $^{13}$CO line emission. For phase 1, we find that the innermost disk region exhibits an $m=1$ spiral in the gas kinematics, which is not seen in the Fourier decomposition of the scattered light (see Fig.~\ref{fig:bhlh_spectrum}) due to the perturbations introduced by the ongoing infall. It is visible in the emission of both isotopologues, indicating that it is not only the result of direct streamer emission. It is instead a result of a kinematic perturbation induced by the streamer. This is apparent because the perturbation in the $^{13}$CO emission spatially matches a perturbation in the $^{12}$CO emission, originating from the upper layers, and this perturbation is connected to the direct emission of the streamer. We also find an $m=1$ spiral in the outer disk regions, which spatially matches one of the arms found in the scattered light (see~Fig.~\ref{fig:bhlh_spirals_rphi}).

We find the most interesting kinematic structure for phase 2. Here, we see strong residual motions in the inner disk that coincide with the location of the brightest scattered-light structures. The inner disk spirals exhibit an $m=2$ pattern in the CO emission, which is especially visible for $^{13}$CO. The outer disk regions show residual motions that are akin to an $m=1$ spiral. These findings are the direct opposite of the apparent scattered-light structures, where the Fourier decomposition shows a strong $m=1$ mode is present in the inner disk, and a strong $m=2$ mode is present in the outer disk (see Fig.~\ref{fig:bhlh_spectrum}, where distinct spiral arms are marked with arrows). A determination of the number of spiral arms by eye suggests that there is an $m=1$ spiral in the outer disk, or an $m=2$ spiral where one arm is very dim, which is matched by the kinematic analysis. Considering the CO lines emission, it is easier to discern observable features caused by the streamer from those exhibited by the disk itself due to jumps in the velocity residuals. The emission of molecular lines is also not directly affected by shadowing (but it is affected indirectly via changes in the temperature). This allows us to conclude that streamer contamination and shadowing are likely explanations for the differences between structures seen in the CO emission and the dominant modes of the Fourier-decomposition of the scattered light.

In contrast to the cloudlet capture case, we find no meaningful structure in the CO isotopologues' residual motion if the disk is viewed at a \SI{30}{\degree} inclination, which is shown in Appendix \ref{sec:app_xco_bhl}. However, more careful modeling of the disk motion that is less susceptible to issues related to environmental emission and non-smooth changes in the emission height might allow for the discovery of residual motions in the more inclined case. Including photodissociation would remove emission from the environment and could further improve the analysis of inclined disks. This is subject to future work.

\subsubsection{Perturbation depth}
The depth toward the midplane of the perturbations induced by the infall in this scenario is overall comparable to what we find in the cloudlet capture case (Section~\ref{sec:cldl_depth}), see Fig.~\ref{fig:bhlh_depth}. As expected, the deepest perturbation occurs during phase 2, when material is actively accreting in the form of a streamer, reaching down to $z\approx 3H$ in the outer disk. Differently from the cloudlet capture case, the perturbation depth does not vary greatly in time, because the accretion does not stop during the simulation. The minimal height above the midplane for which the density CV is larger than $0.1$ is shallower than during the cloudlet impact. This indicates that the disk density structure is not noticeably affected by the infall beyond the upper surface layers.

\subsubsection{Gas orbital elements}
\begin{figure*}[htp]
    \centering
    \begin{minipage}[t]{.65\linewidth}
        \vspace*{0pt}
        \includegraphics[width=\linewidth]{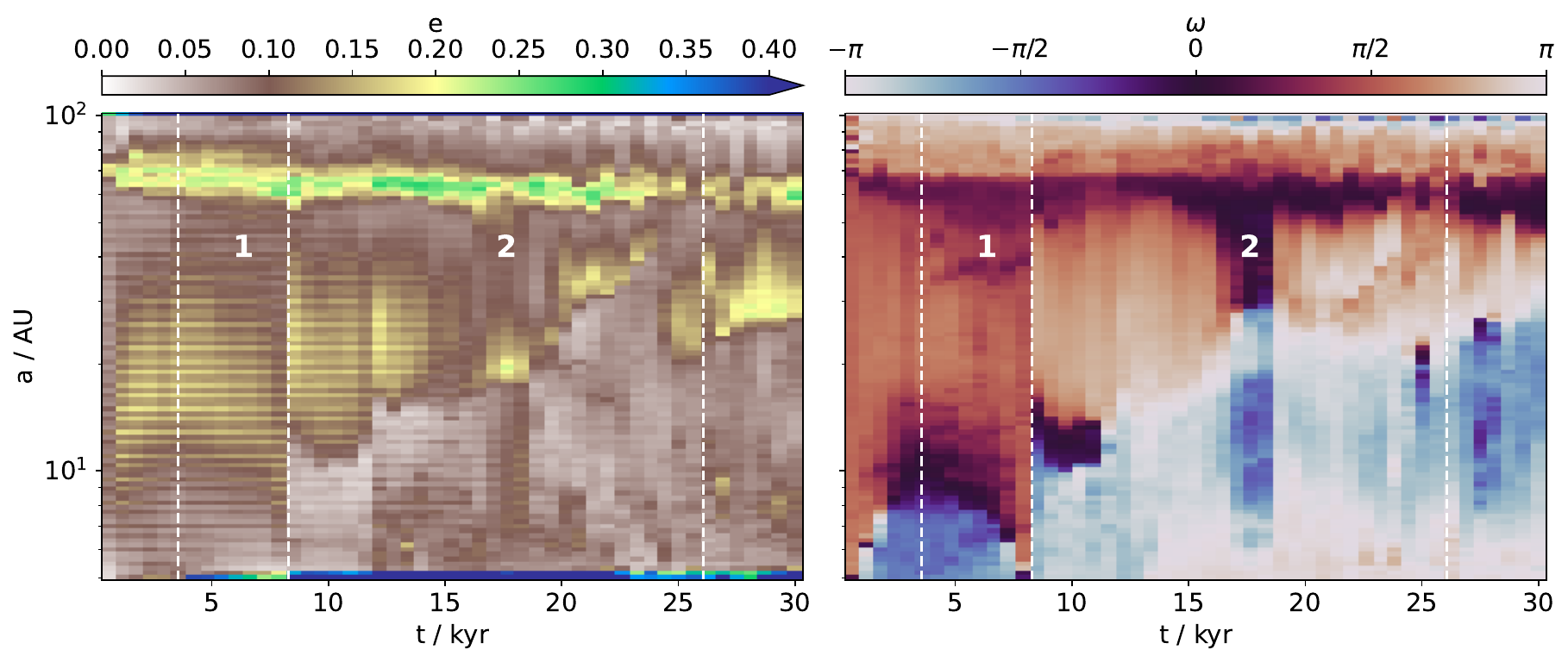}
    \end{minipage}%
    \begin{minipage}[t]{.25\linewidth}
        \vspace*{.5cm}
        \includegraphics[width=\linewidth]{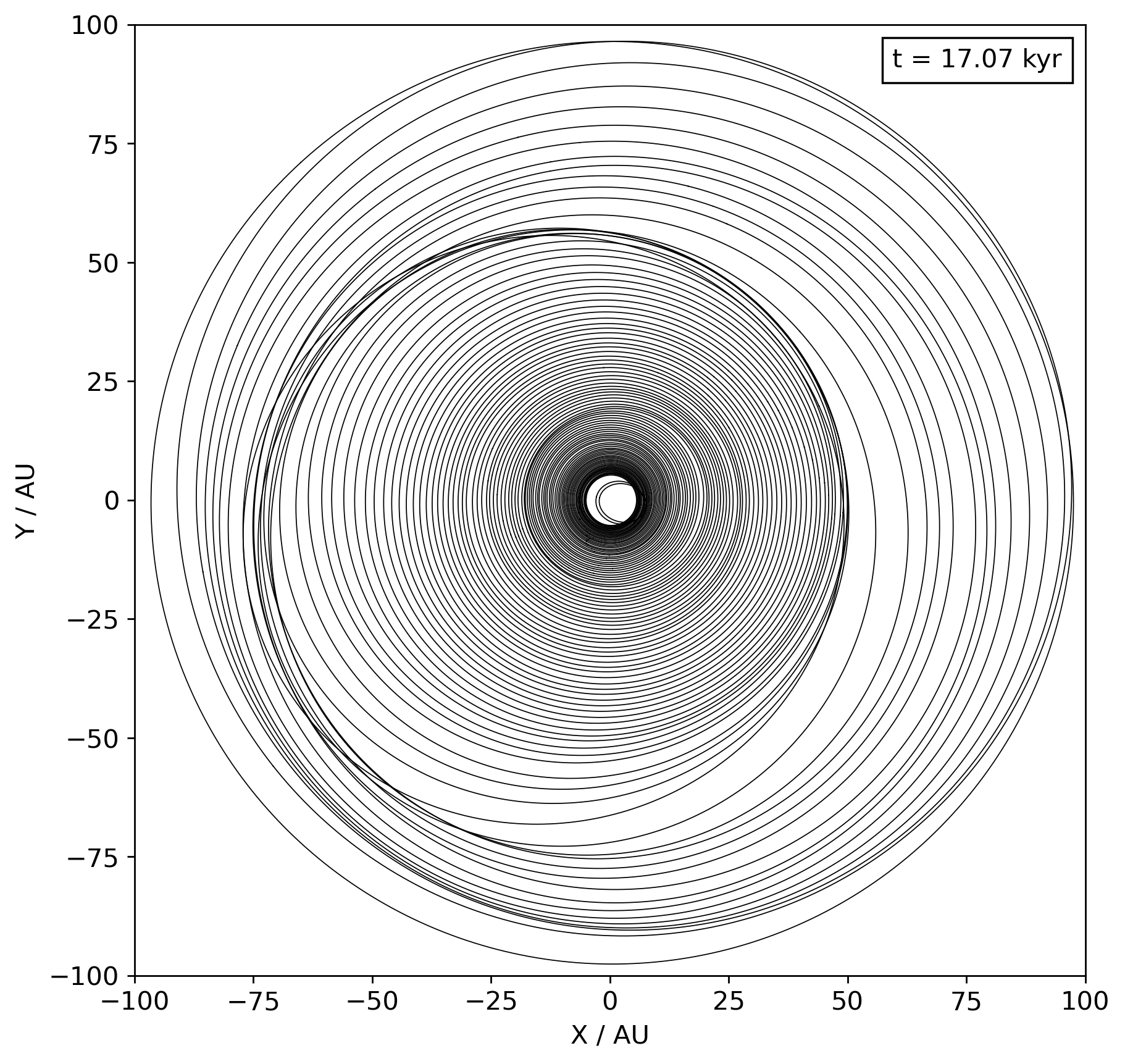}
    \end{minipage}%
    \caption{Same as Fig.~\ref{fig:cldl_orbit}, but for simulation 2.}
    \label{fig:bhlh_orbit}
\end{figure*}%
We analyze the orbital elements in the same way as in the cloudlet capture case. The accretion of material leads to an excitation of the eccentricity in the surface layers, and a simultaneous shift in the argument of periastron, shown in the left and middle panel in Fig.~\ref{fig:bhlh_orbit}. There are several key differences to the excitation of eccentricity in the cloudlet capture scenario. First, the maximal excitation is slightly lower, reaching $e\sim 0.25$, present during phase 2 when the streamer is active. Second, the semi-major axes of the excited gas are limited to a smaller range of values, close to $60$ -- $\SI{70}{\astronomicalunit}$. Lastly and most importantly, the alignment of the elliptical orbits caused by the shift in the argument of periastron creates an m=1 spiral pattern, shown in the right panel in Fig.~\ref{fig:bhlh_orbit}, similar to the findings by \citet{calcino2025a}.

Even though we find a similar alignment of the gas orbits, we find the resulting pattern to be unrelated to the spiral patterns that arise in the scattered light synthetic observation. While the strong $m=1$ mode in the inner disk (see Fig.~\ref{fig:bhlh_spectrum}) may initially suggest the presence of a spiral created through this mechanism, its shape and location do not match the scattered light images found in Fig.~\ref{fig:bhlh_spirals_rphi}. Furthermore, the ellipse pattern is already present during phase 1, where the $m=1$ component of the Fourier decomposition is weak in both the inner and outer disk, and where no visual match to the pattern exists. The disk eccentricity could, in principle, also be responsible for the spiral patterns found in the residual motions, motivated by the idea that gas in an eccentric disk exhibits vertical motion due to a ``breathing mode'' \citep{ragusa2024}. Indeed, in the CO residual motion of the two considered isotopologues, various $m=1$ patterns arise, but they are of a different shape and at a different location. We therefore conclude that, while the orbits are affected in the same way a found by \citet{calcino2025a}, the spiral patterns we find in the scattered light and residual motions are a direct consequence of the late infall, rather than an alignment of eccentric orbits. The disturbances at the top layers of the disk due to the infall act to conceal any structures that the orbital shift might create, and the lower disk layers traced by the less abundant isotopologues are not sufficiently perturbed.

\subsection{Lower disk mass}
In previous sections, we found that the induced spiral structures are mainly caused by the disturbances in velocity and density in the upper layers of the disk. The reason for the shallow influence of the infall on the disk is the comparatively low mass of accreted material compared to the disk mass. We therefore expect the mass of the disk itself to have a major impact on our results, as a lower-mass disk would be subject to deeper perturbations. In this subsection, we investigate both infall scenarios, a cloudlet capture scenario and gas accretion from a turbulent medium, with a reduced disk mass by a factor of 10. We summarize the most important differences in the following. A more detailed analysis is given in Appendix \ref{sec:app_lmd}.

\begin{figure}[htp]
    \centering\includegraphics[width=\linewidth]{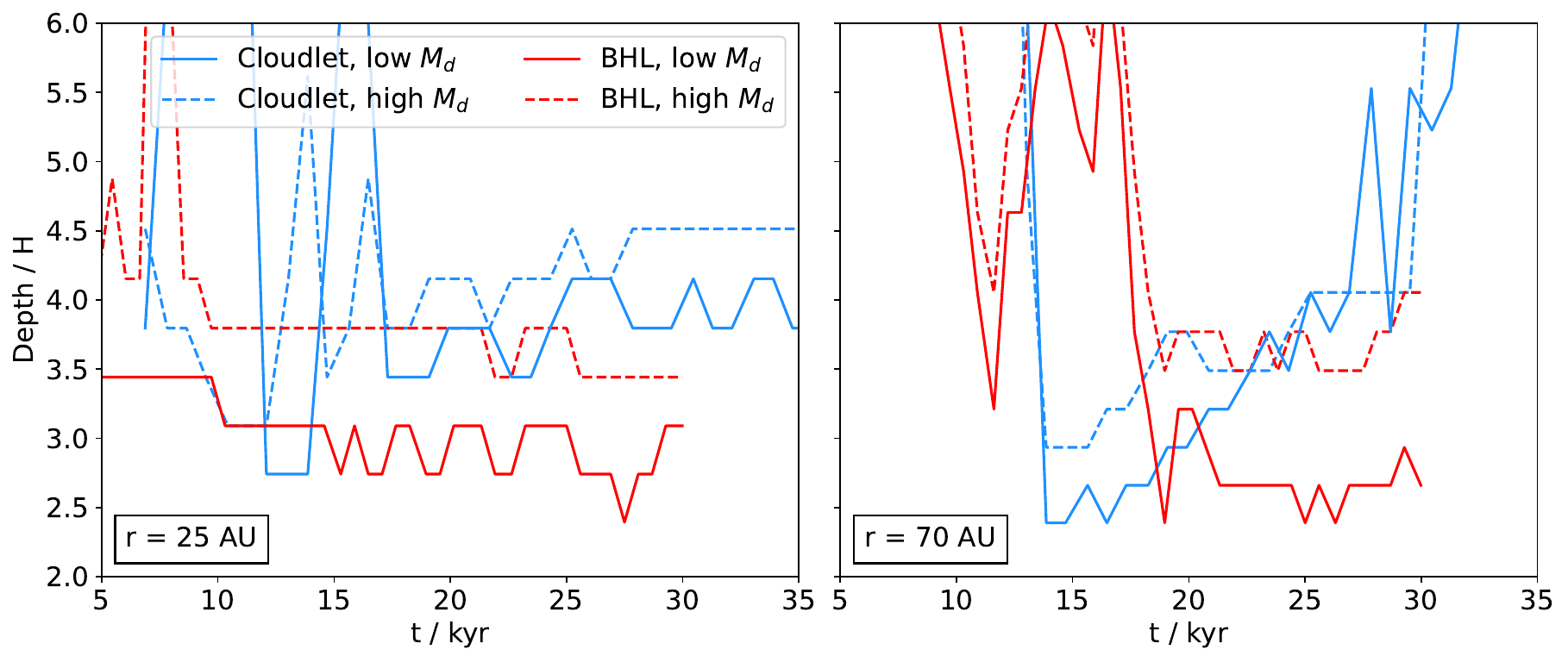}
    \caption{Perturbation depth in units of the pressure scale height as a function of time for all four simulations. The left panel shows the depth for the inner disk ($r=\SI{25}{\astronomicalunit}$), and the right panel shows the depth for the outer disk ($r=\SI{70}{\astronomicalunit}$). Dashed lines represent simulations with regular disk mass, solid lines ones with low disk mass. The blue lines describe cloudlet capture simulations, whereas the red ones describe accretion in a turbulent medium.}
    \label{fig:comp_depth}
\end{figure}%
As expected, the disks are more deeply perturbed if the disk mass is lower, which is more pronounced for the turbulent accretion scenarios. This is shown in Fig.~\ref{fig:comp_depth}, where we show the lowest value of $z/H$ for which the CV of the density is at least $0.5$. The figure can also act to quantify the results discussed in previous sections: The perturbations by BHL accretion are deeper at later times, but shallower than the perturbation by the cloudlet capture during the time when the cloudlet encounters the disk. This result generally holds true for the lower disk mass case, but here, the BHL perturbation is deeper than the cloudlet capture one at earlier times. Specifically, it occurs at $t\sim\SI{15}{\kilo\year}$ in the inner disk and $t\sim\SI{20}{\kilo\year}$ in the outer disk for the lower disk mass, but at $t\sim\SI{20}{\kilo\year}$ in the inner and $t\sim\SI{25}{\kilo\year}$ in the outer disk for the regular disk mass cases. For the lower disk mass, the deepest perturbation is 2.5 pressure scale heights for BHL accretion for both the inner and outer disk, although this occurs at different times for the different disk regions. In the case of cloudlet capture, the disk with lower mass is perturbed down to 3 pressure scale heights in the inner, and 2.5 pressure scale heights in the outer disk.

\begin{figure*}[htp]
    \centering\includegraphics[width=.75\textwidth]{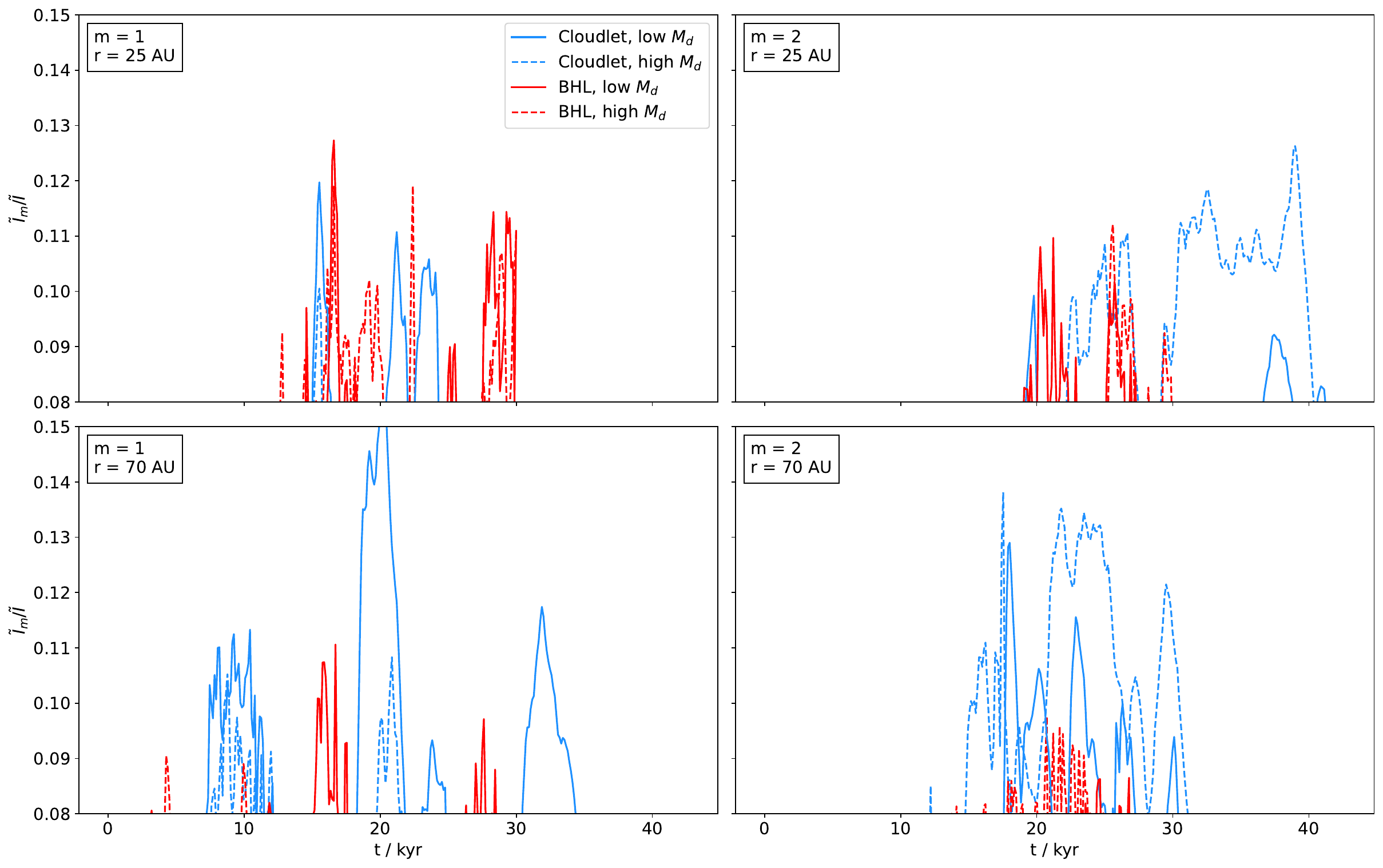}
    \caption{Normalized, Fourier-transformed intensity of the $m=1$ (left panels) and $m=2$ (right panels) modes of the polarized scattered light intensity of all four simulations. The top panels show the intensity in the inner disk ($r=\SI{25}{\astronomicalunit}$), whereas the bottom panels show the intensity in the outer disk ($r=\SI{70}{\astronomicalunit}$).}
    \label{fig:comp_spectrum}
\end{figure*}%
A natural consequence of the change in perturbation depth is a corresponding change to the arising spiral structures in the polarized scattered-light intensity, especially for the low spiral modes. We provide a comparison for all four simulations in Fig.~\ref{fig:comp_spectrum}. Least impacted by the change in disk mass is the relative amplitude of the $m=1$ mode in the inner disk. Here, the amplitude is diminished compared to the regular mass disk around $t\sim\SI{20}{\kilo\year}$ for the BHL accretion case, but enhanced around the same time for the cloudlet capture case. More significant differences occur in the outer disk. Here, the amplitude of the $m=1$ mode is strongly enhanced for the cloudlet capture case at $t\sim\SI{20}{\kilo\year}$, and again at $t\sim\SI{30}{\kilo\year}$. The increase is less severe, but present for the turbulent medium accretion simulation, at $t\sim\SI{15}{\kilo\year}$.

We find that for the $m=2$ mode, a decrease in disk mass is detrimental to the relative amplitude. This is especially true for the cloudlet capture case, where the mode disappears fully in the inner disk. In the outer disk, while not disappearing fully, it is strongly decreased for both the cloudlet capture case and the scenario where gas is accreted from the turbulent medium.

\section{Discussion}\label{sec:discussion}

\subsection{Spiral formation mechanism}
In this work, we show that both considered infall mechanisms, that is, cloudlet capture and BHL accretion, can cause spiral structures, and that the $m=2$ mode dominates the arising pattern. This result deviates from the results of previous studies, that mainly found the $m=1$ mode to dominate in the case of asymmetric accretion \citep{hennebelle2017,calcino2025a}. A more detailed analysis of the physical origin of the spiral structures is subject of future work, but in the following, we discuss the main potential spiral formation mechanism in our simulations, and the main differences to previous work.

Without consideration for secondary mechanisms, the effects of arbitrary perturbations to any protoplanetary disk quantity on its dynamics can be studied by performing a decomposition into a series, for example, azimuthal Fourier modes or vertical Hermite modes (see, e.g., \citealt{zhang2025}). A perturbation whose decomposition is dominated by an $m=1$ Fourier mode would then be associated with creating an $m=1$ spiral pattern. To first order, this is the reason for the creation of such a pattern in previous works, where the infall is highly asymmetric. The infall originates from one direction and interacts with the side of the disk facing that direction, either fully in the case of extended infall, or in a spatially confined way if the infall takes the form of a streamer. Other mechanisms can create $m=1$ spirals, for example, a disk warp \citep{winter2025} or the excitation of disk eccentricity and a twist \citep{calcino2025a}. However, we find that these mechanisms are not responsible for the spirals arising in our simulations.

Such a strong asymmetry in the interaction with the infalling material does not exist in our simulations. This is a consequence of the fallback of material in both the cloudlet and BHL case (see HD25). In the simulations of cloudlet capture, the expanded cloudlet interacts with the disk predominantly from one side during the initial encounter, and the newly bound, initially overshooting material falling back interacts with the disk predominantly from the other side. This double-sided interaction has a strong $m=2$ mode in its Fourier decomposition, leading to $m=2$ spirals. These spirals are not identifiable in synthetic observations during the entire interaction. The reason for this is that the disk surface is perturbed by the accretion, in the form of sound- and shockwaves. This is especially relevant during the initial interaction between the cloudlet and the disk, and during the period where the fallback streamer is active. After these strong interactions have subsided, the spirals are more clearly visible; they are still caused by font- and backside interactions with the leftover cloudlet gas, but the accretion rate is lower and corresponding surface perturbations are less relevant. The two-sided interaction being the main driver of the $m=2$ spiral morphology also offers an explanation for the result that the spiral arms of the cloudlet-induced $m=2$ spirals are almost stationary: The spirals are launched from the interaction sites between disk and infall, which do not change drastically over the course of the simulations. This matches the previous findings by \citet{calcino2025a} and \citet{hennebelle2017}.

The mechanism forming $m=2$ spirals in the BHL accretion simulations is similar in concept, but the details are different. The two-sided nature of the interaction is given by the systemic flow on one side, and the fallback of material from the accretion tail on the other side. While a turbulent medium is necessary to create fallback streamers (see HD25), turbulence is not a required ingredient to form $m=2$ spirals, as the two-sided interaction arises due to the fallback of material in the accretion tail. If turbulence is included in the consideration, the $m=2$ spiral formation mechanism does not change, because the deviations it introduces to the density and velocity fields of the systemic ISM flow are small and do not have a significant impact on the fallback mechanism. Therefore, this scenario is best compared to the \texttt{ASNR2} case of \citet{hennebelle2017}, where inflow is asymmetric but non-rotating. While not discussed in the original work, this simulation also shows a weak $m=2$ spiral pattern. In our study, we do not perform a pattern speed analysis for the BHL accretion simulations, because the surface is too strongly perturbed in the scattered light to allow for a proper characterization of the spiral location. However, given that the spirals presumably form through the same mechanism, we expect the spirals to also be almost stationary in the BHL accretion case.

In summary, the mechanism forming $m=2$ spirals in our simulations is the two-sided interaction of infalling material with the disk, caused by the fallback of newly bound material. The essence of this mechanism is the same for both accretion mechanisms, although their observational implications are different. In the cloudlet capture case, the main interaction leads to stronger perturbations, but the corresponding spiral patterns are difficult to discover due to the corresponding surface perturbations. The interaction with the leftover gas after the main encounter is favorable for the detection of spirals, because the accretion occurs without creating a fallback streamer, but it is limited in time. On the other hand, BHL accretion creates more persistent interactions, solving the issue of a small interaction time compared to the disk lifetime. At the same time, the persistent interactions lead to the continuous creation of fallback streamers, and the corresponding constant surface perturbation make the detection of spirals more challenging in observations.

\subsection{Implications for observational interpretation of spirals}
We find a large variety of spirals arising as the result of surface-level disturbances solely caused by late infall. In particular, we find that prominent $m=2$ spirals can arise in scattered light, and $m=1$ spirals in the CO residual motions, which are unrelated to other mechanisms commonly invoked to form spirals. The infall that causes these structures also does not necessarily need to be visible in the form of a streamer. In fact, the most prominent scattered light spiral structure arises in Fig. \ref{fig:cldl_spirals_rphi}, at a time when only a faint remnant of the original cloudlet is accreted after the main infall event, with an active fallback streamer, has passed. With this in mind, special care should be taken when inferring disk conditions or planet candidates from the existence of spiral structures alone. If no additional information, such as a considerable pattern speed or perturbations in deeper layers, is available that can exclude a late-infall event causing spirals, an observed spiral might be mistaken for evidence of a perturber, GI, or a warped disk that might not reflect the physical reality. For example, a warped and twisted disk was recently found to be a possible explanation for the spirals in the $^{12}$CO line emission in MWC 758 \citep{winter2025}; however, none of our spiral structure, neither in scattered light, nor in CO, is related to a warp structure (see Appendix \ref{sec:app_warp}).

Deep perturbations could act as evidence for the formation of spirals through other mechanisms, because late infall does not drive them directly unless a lot of mass is accreted (see Appendix \ref{sec:app_lmd}). It remains to be investigated to which extent this statement holds true when considering that the infall could trigger secondary mechanism, like the Rossby wave instability \citep{bae2015,kuznetsova2022}, or the excitation of turbulence (e.g., \citealt{fs2003}). It is currently unclear if the Rossby wave instability could be triggered by the infall if mass is not directly deposited at the midplane, as 3D studies of it assume hydrostatic equilibrium (e.g., \citealt{meheut2012}), which does not hold in this case.

Based on the patterns we see in our simulations, we predict that structures like in SU Aur \citep{ginski2021} should be most common for disks under the influence of late infall: A clearly visible streamer with flocculent spiral structures. Especially in the inner disk, we find that the perturbation caused by streamers prevent the formation of spirals with strong low-numbered Fourier modes, so that we would not expect to see such spirals at the same time as a streamer. This conclusion holds for observations of scattered light; even in the abundant CO isotopologues, low-m spirals can exist during active infall streamers. In general, the spiral morphology in scattered light and CO line emission could be used to infer information about the environmental conditions causing streamers of material, by distinguishing between accretion from a turbulent ISM or accretion caused by a single encounter with a cloudlet-like overdensity.

The spiral patterns that we discover in the CO line emission kinematic residuals are the clearest when the disk is viewed almost face-on, with only a small deviation from a perfect \SI{90}{\degree} angle being introduced by the infall. This indicates that the motions causing the spiral patterns have a strong vertical component. We find that, during the accretion of the cloudlet remnant, the spirals are not visible if the disk is viewed at an inclination of $i=\SI{30}{\degree}$. They are only visible while the streamer is active in the cloudlet capture case. However, this could be related to difficulties in fitting a Keplerian model to the emission, caused by the abundant emission of the surrounding material. More detailed modeling, or the consideration of the photodissociation of CO which reduces emission in the surrounding cloud, could allow for the detection of spirals also in the inclined case. If that is indeed the case, we would expect the perturbations at the surface level, caused by the turbulent flow, to differ between the front and back emission surfaces, which could act as another way to distinguish infall-related perturbations from those created by other mechanisms.

As discussed in the previous sections, there are two key factors that allow us to distinguish the spirals caused by late infall from spirals caused by other mechanisms: The pattern speed and the dependence of the structures on the height above the midplane. In some cases, the shape of the spirals themselves can also aid in the distinction. For example, depending on the cooling rate, GI causes flocculent structure in the outer, and spirals with fewer arms in the inner disk \citep{rowther2024}. This is the direct opposite of infall-induced spirals, where flocculent structure is found in the inner disk, and $m=2$ spirals are found in the outer disk.

\subsection{Implications for disk dynamics and planet formation}
The presence of spiral substructures in the disk raises the question whether they could aid planet formation. On the one hand, infall generally delivers new solids that can aid, or even rejuvenate, planet formation. In addition, spirals can accumulate dust and create density enhancements that would be beneficial, for example for the formation of planetesimals or for retaining solids for pebble accretion. For this to be effective, trapping needs to occur at the disk midplane, but we find that the midplane is not directly affected by the infall; the spiral structure does not penetrate deep enough into more massive disks. For these disks, that is, for $M_d\gtrsim\SI{0.05}{\solarmass}$, we therefore conclude that the induced spirals do not influence planet formation neither positively, nor negatively.

This finding does not come without a model-based caveat. The presence of a strongly turbulent disk surface, excited by the infall, could drive weak turbulence in the midplane (e.g., \citealt{fs2003}), which would be missed in our simulations. Numerical viscosity introduced by the misalignment between the gas velocity and the cell surfaces is not a significant problem due to our choice of a spherical grid, but numerical diffusion might play a significant role. But even if that was fully negligible, low levels of turbulence could not be resolved in our simulations, because the eddy scale, $\sqrt{\alpha}H$, is smaller than the cell size. In fact, resolving the excitation of turbulence with $\alpha=\num{e-2}$ requires a cell size of $0.1H$, or 10 cells per scale height, which is not reached inside the disk, but even turbulence below this value could influence planet formation. All in all, while our results indicate that the midplane is not directly perturbed by the infall, our setup does not allow us to make a definitive statement about whether planet formation could be aided by the secondary excitement of turbulence via the turbulent upper layers.

This result strongly depends on the ratio between the accreted mass through infall and the mass of the already present protoplanetary disk. We find that, for a lower initial disk mass but the same accreted mass, the perturbations reach deeper into the disk (see Appendix \ref{sec:app_lmd}). If the infall perturbs the disk deep enough to impact the midplane structure, dust traps can be created, impacting planet formation. This would be more likely to occur at later stages of the disk, when most of the mass has been depleted. Alternatively, a higher infall rate onto a disk of unchanged mass would also affect layers close to the midplane, but might lead to the disk and surrounding cloud being interpreted as a Class I source. This suggests that late infall may be especially important to consider toward the end of the protoplanetary disk's lifetime.

\subsection{Model caveats}
Our simulations only consider the evolution of the gas in the disk and its surroundings, and we make no attempt to model the transport or evolution of any solids. This is motivated by the computational cost that would be introduced by integrating additional components, especially if multiple grain sizes and grain growth are to be treated. For the large-scale streamers themselves, it is reasonable to assume that the dust is small and therefore well-coupled, so that an inclusion of the dust is not very impactful. While assuming perfectly mixed and small grains is also suitable for the consideration of scattered light and CO line emission images at sufficient emission heights, modeling of the dust size distribution, including coagulation, and dust-gas interactions is necessary for a more detailed investigation of the impact that late infall has on dust transport, dynamics, and planet formation. Even though we only find surface-level perturbations in the gas, the growth, settling, drift, and vertical mixing could, in principle, connect the perturbed surface with the seemingly unperturbed midplane and impact planet formation directly.

We modeled the BHL accretion by introducing compressible turbulence to the large-scale environment around the disk as an initial condition, but without dynamically driving the turbulence. As a result, the turbulence decays over time in our model, whereas in reality, the strength of the turbulence at a given time is a reflection of the environmental conditions in the region the disk moves through. The decay is more significant for the smaller scales, as scales smaller than the cell size cannot be resolved, and with our choice of spherical coordinates with log-radial spacing, the cell size increases with increasing distance from the star. Flow that originates from these further-out regions, reaching the disk at later times, has a turbulent power spectrum that is suppressed at small scales (larger $k$ values). We do not expect that this decay has a significant impact on our results, but we note that the emergence of spirals that have low-m Fourier modes with a high amplitude could be related to this artificial suppression of small-scale turbulence at that time. As we utilize a simple isothermal temperature model in our simulations, we also do not treat the heating that would occur as a result of the shocks and turbulent compression in the environment. The former is likely to be result in substantial heating as streamers hit the disk surface \citep{vangelder2021}. In the future, it should be investigated how this heating affects the spiral structures.

We did not consider the effects of freeze-out and photo-dissociation when computing the CO abundances used for the creation of the synthetic CO line emission images. While we do not expect these effects to have a significant impact on the abundances inside the disk and at its surface, the resulting optical depth of the cloud surrounding the disk is likely overestimated as a result. As a result, it has a stronger effect on the \texttt{discminer} Keplerian fit, and fits performed on a real observation might have more success with discovering residual motions without introducing emission-height related model residuals. As we only resolve the disk with a few cells per pressure scale height in the vertical direction, we might also incur inaccuracies in the effective height above the midplane a particular CO isotopologue is emitting from.

\section{Conclusions}
We performed 3D hydrodynamics simulations and radiative transfer modeling to investigate the emergence of spiral structures as a result of late infall. For our investigation, we used the simulation setups leading to streamers from \citet{hd25}, considering both the capture of a gaseous cloudlet and the accretion of gas from a turbulent medium. We draw the following conclusions.
\begin{itemize}
    \item In the cloudlet capture scenario and considering the scattered light emission, the inner disk is strongly perturbed once the cloudlet hits the disk and a streamer forms, leading to flocculent structure. In contrast, $m=2$ spirals can form in the outer disk at this time, which only occurs in the inner disk after the initial streamer has subsided. This scenario generally favors the formation of well-defined spiral arms after the infall has largely subsided.
    \item The pattern speed of the $m=2$ spiral during the remnant accretion after cloudlet capture is low, ranging between $\dot{\phi}=\SI{0.05}{\per\kilo\year}$ ($r_\mathrm{corot}=\SI{3150}{\astronomicalunit}$) and $\dot{\phi}=\SI{0.11}{\per\kilo\year}$ ($r_\mathrm{corot}=\SI{1460}{\astronomicalunit}$), depending on the considered radius and spiral arm.
    \item In the scenario where the disk accretes mass from a turbulent medium, where the infall rate fluctuates but does not fully subside, the formation of surface-level $m=2$ spirals, visible in scattered light, is generally not favored. Here, the disk shows more flocculent structures during the simulation, but before the first streamer arises, an $m=2$ spiral can form.
    \item Spiral structures can also be seen in $^{12}$CO line emission through residuals from Keplerian motion, but they do not match the scattered light emission patterns. They are best seen when the disk is viewed face-on, suggesting that they are caused by vertical motion. The amplitude of their individual angular modes also differ from the scattered light case.
    \item The perturbations induced by the infall are mostly limited to the upper layers of the disks, to about four pressure scale heights. This is directly related to the smaller residual motions found in the emission of the more optically thin $^{13}$CO.
    \item While changes to the gas orbital elements can be induced for some time in the upper layers, the $m=1$ spiral pattern induced for the turbulent medium accretion case is not visible, neither in scattered light, nor in molecular line emission.
    \item Perturbations reach deeper into a disk if it has low mass compared to the infall, which is expected. This influences the spirals that emerge in scattered light in many aspects; most notably, it disfavors the emergence of an $m=2$ spiral during remnant accretion.
\end{itemize}
Our results demonstrate the observed spiral structures do not necessarily need to be induced by a clear perturber, even if the spiral arms are very well-defined. Instead, they emerge naturally as a result of infall. Despite this, they can be discerned from spirals formed through other mechanisms through their low pattern speed. More generally, the spiral perturbations do not reach the midplane of the disk, unless its mass is low compared to the accreted mass, which is the most relevant location for planet formation. Spirals formed solely through infall, without considering any additional mechanisms they could trigger, are therefore not expected to greatly impact planet formation beyond the delivery of new material, especially at the beginning of the Class II stage.

\begin{acknowledgements}
We thank the anonymous referee for suggestions that helped improve the manuscript. We thank Haochang Jiang and Josh Calcino for stimulating discussions. L.-A. H. acknowledges funding by the DFG via the Heidelberg Cluster of Excellence STRUCTURES in the framework of Germany's Excellence Strategy (grant EXC-2181/1 -- 390900948). We acknowledge support by the High Performance and Cloud Computing Group at the Zentrum für Datenverarbeitung of the University of Tübingen, the state of Baden-Württemberg through bwHPC and the German Research Foundation (DFG) through grants INST 35/1134-1 FUGG, 35/1597-1 FUGG and 37/935-1 FUGG.
\end{acknowledgements}
\bibliography{references}
\begin{appendix}
\onecolumn
\section{Additional figures for turbulent medium accretion (high disk mass)}\label{sec:app_xco_bhl}
\begin{figure*}[htp]
    \centering\includegraphics[width=\linewidth]{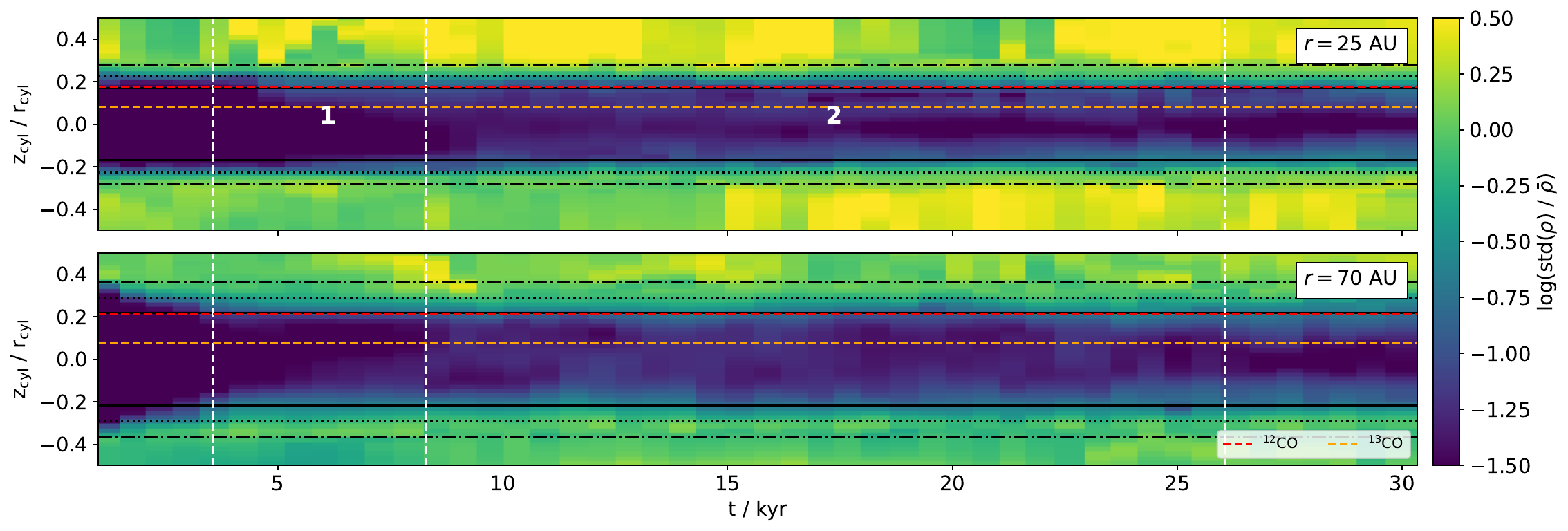}
    \caption{Same as Fig.~\ref{fig:cldl_depth}, but for simulation 2.}
    \label{fig:bhlh_depth}
\end{figure*}%
\begin{figure*}[htp]
    \centering\includegraphics[width=.45\linewidth]{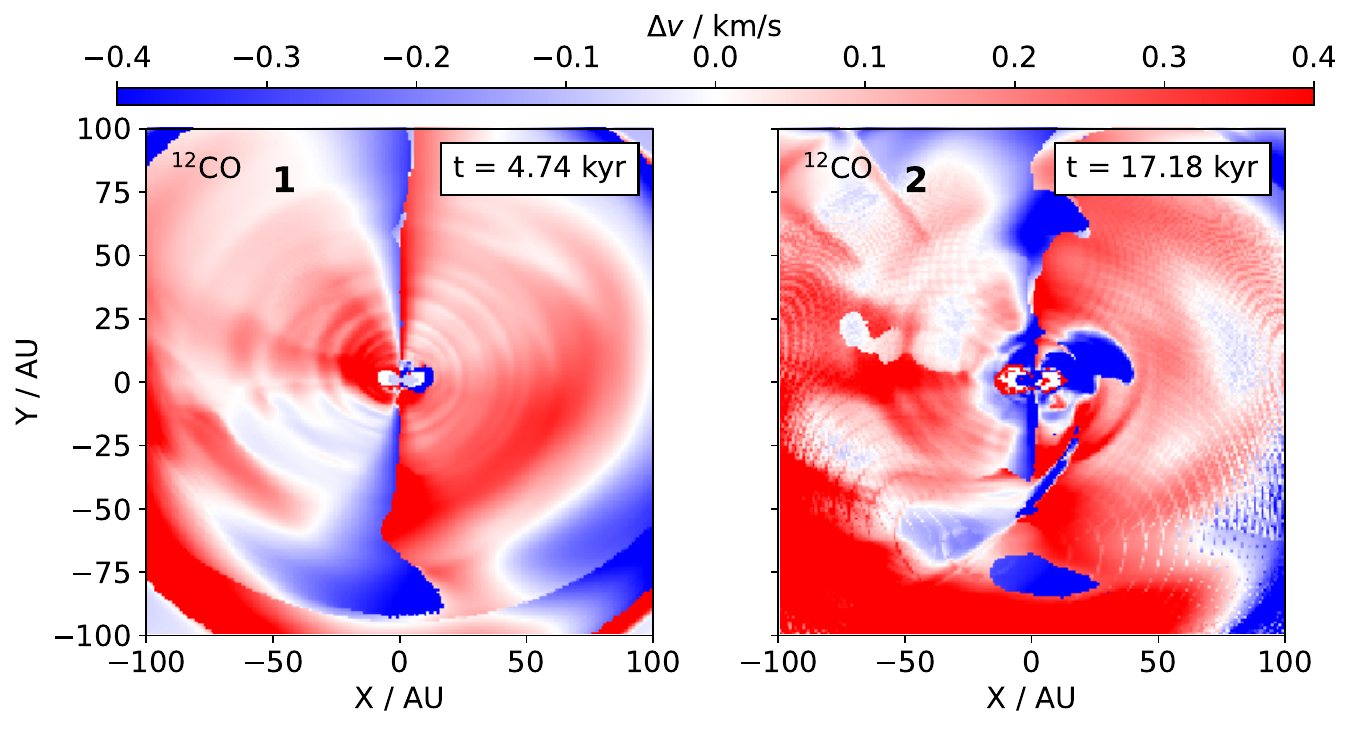}\includegraphics[width=.45\linewidth]{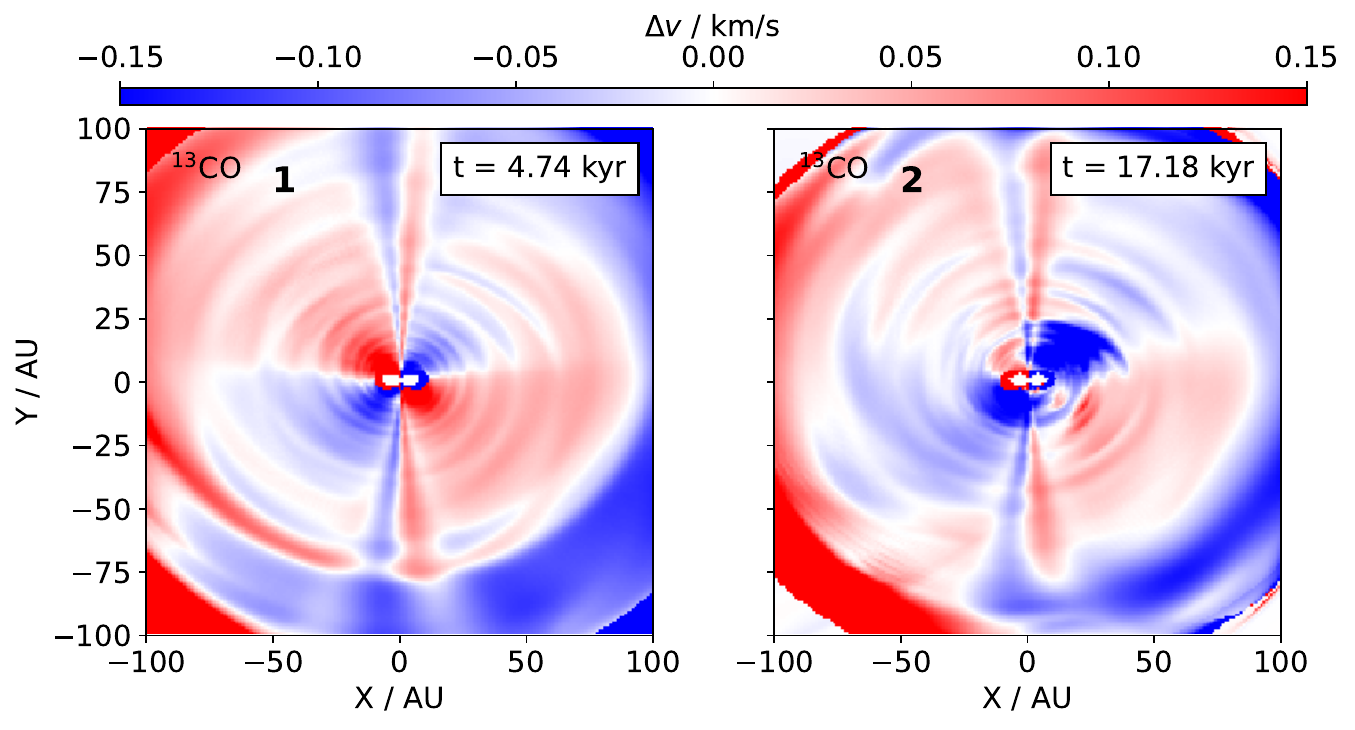}
    \caption{Residual motion found in CO isotopologue line emission for simulation 2 at an inclination of $i=\SI{30}{\degree}$. Left: Same as bottom row of Fig.~\ref{fig:cldl_co}. Right: Same as bottom row of Fig.~\ref{fig:cldl_13co}.}
    \label{fig:bhlh_xco_rest}
\end{figure*}%
In this Appendix, we present additional figures for simulation 2, which were not discussed in Section~\ref{sec:bhlh}. In Fig.~\ref{fig:bhlh_depth}, the depth of the perturbations induced by the BHL accretion is shown. Similar to the cloudlet capture case, the perturbations do not reach deeper than three pressure scale heights in the inner disk, and only does so very briefly for the outer disk. This allows us to conclude that, for a disk mass of $M_d=\SI{0.05}{\solarmass}$, any discovered substructure is limited to the surface for both infall mechanisms, even when assuming initial conditions or infall rates that are on the higher end of realistic values.

In Fig.~\ref{fig:bhlh_xco_rest}, we present the residuals of the moment 1 line emission of $^{12}$CO and $^{13}$CO for an inclination of $i=\SI{30}{\degree}$. Unlike the face-on case, no considerable substructure can be seen. We note that, due to the optical thickness of the surrounding turbulent medium and the strong disturbance of the disk kinematics in the upper disk layers, fitting a Keplerian model becomes increasingly difficult within the capabilities of the discminer code, creating severe non-physical residuals, not allowing any conclusions to be made for an inclination of $i=\SI{30}{\degree}$. In addition, a straight line of residuals can be seen at $x=0$. This is caused by missing emission in the \SI{0}{\kilo\meter\per\second} velocity channel in the synthetic observations, which is the result of the turbulent medium surrounding the disk. Due to the simulation setup, it has no velocity toward the line of sight, so that emission in that channel is dominated by the background emission. A different camera angle would shift this strong residual to a different channel, but would not fix the issue. In reality, the freeze-out of CO might make the line emission less optically thick than we assume (see Section~\ref{sec:discussion}). For $^{13}$CO, the surrounding medium does not obscure the disk as strongly.
\FloatBarrier
\section{Additional figures for simulations with low disk mass}\label{sec:app_lmd}
\begin{figure*}[htp]
    \centering\includegraphics[width=.45\textwidth]{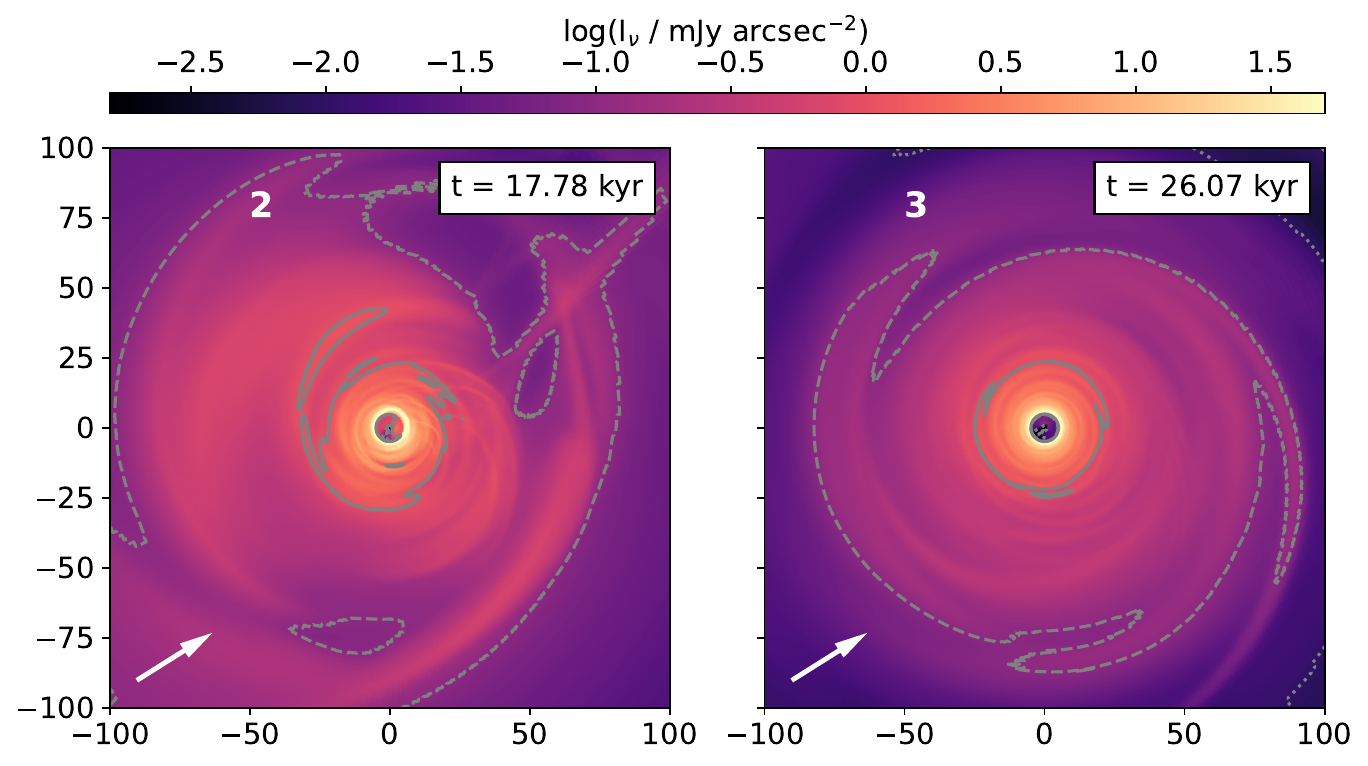}\includegraphics[width=.45\linewidth]{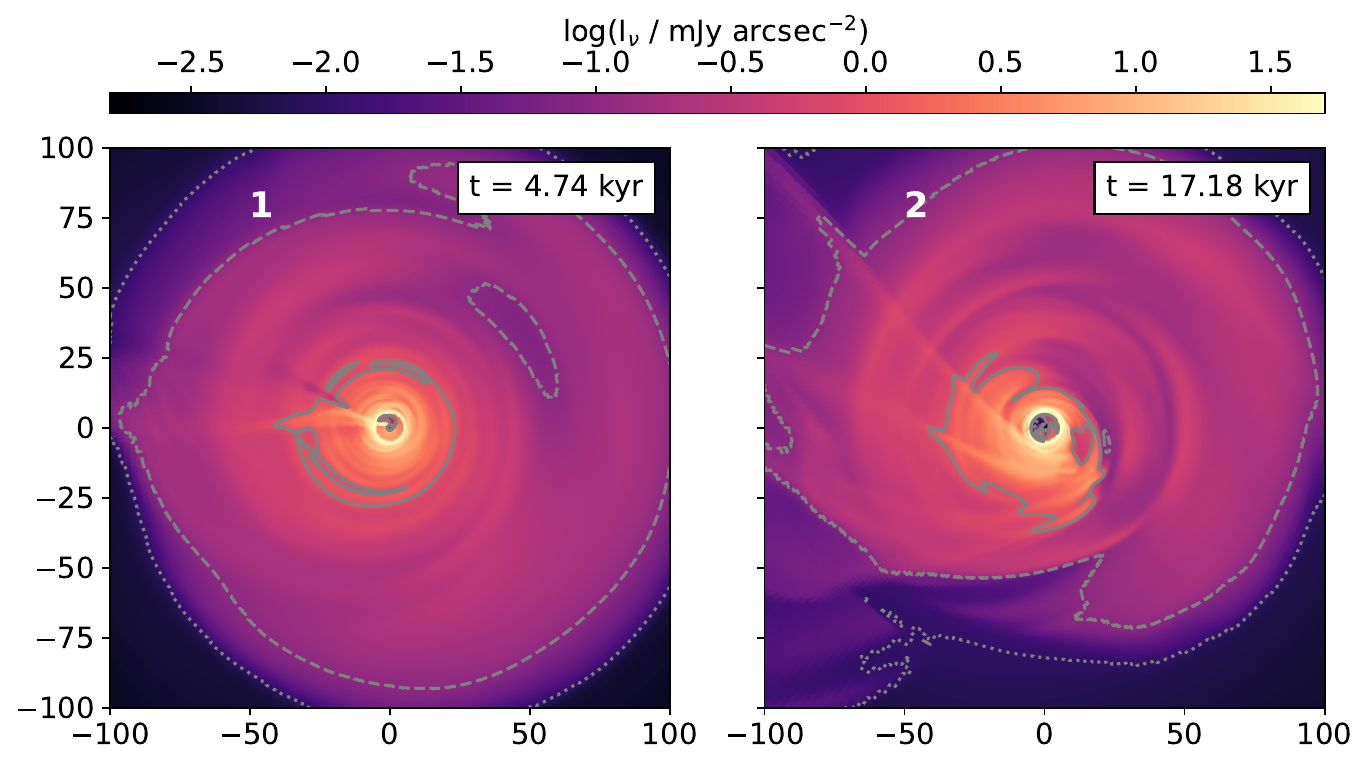}
    \caption{Synthetic scattered light observations of structures found in the simulations with a lower disk mass. Left: Same as zoomed-in area of panels 2 and 3 in Fig.~\ref{fig:cldl_spirals_rphi}, but for simulation 3. Right: Same as zoomed-in areas in Fig.~\ref{fig:bhlh_spirals_rphi}, but for simulation 4.}
    \label{fig:low_spirals_rphi}
\end{figure*}%
\begin{figure*}[htp]
    \centering\includegraphics[width=.45\textwidth]{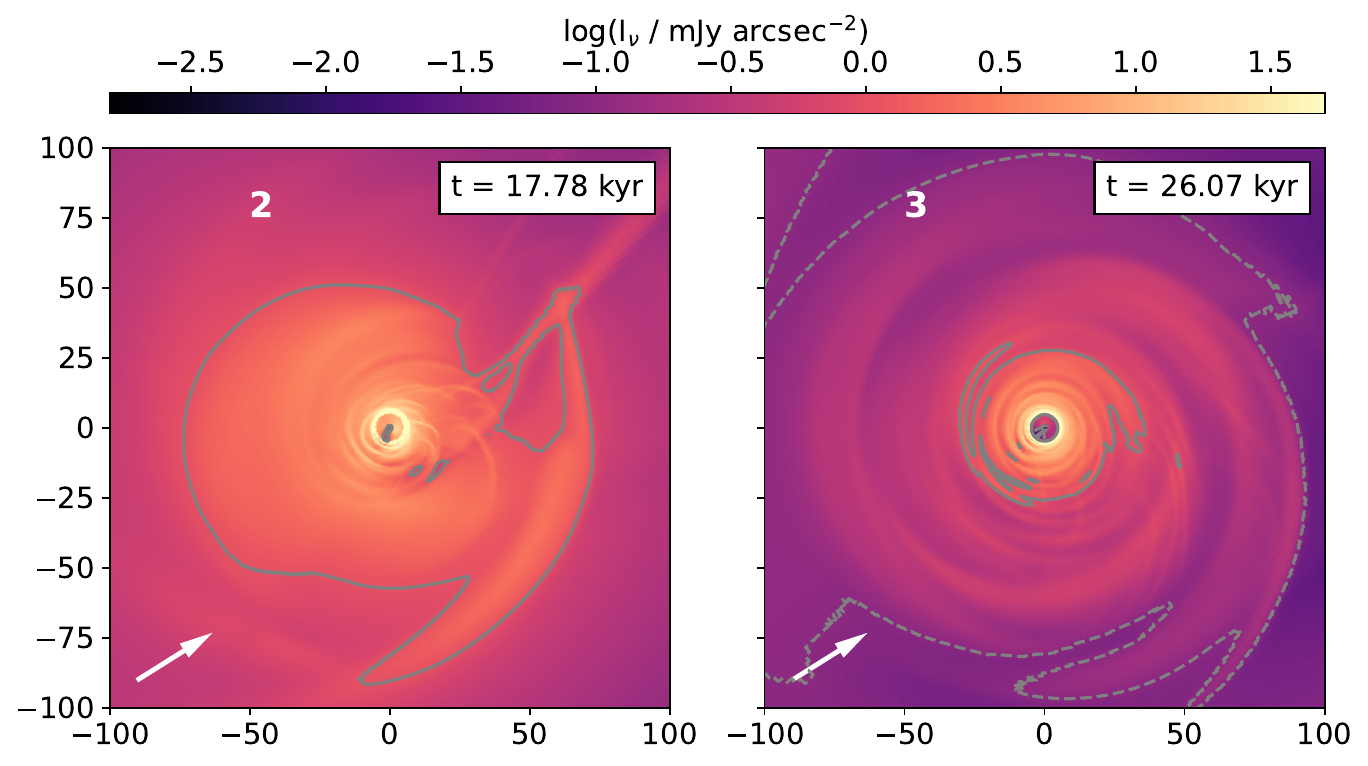}\includegraphics[width=.45\linewidth]{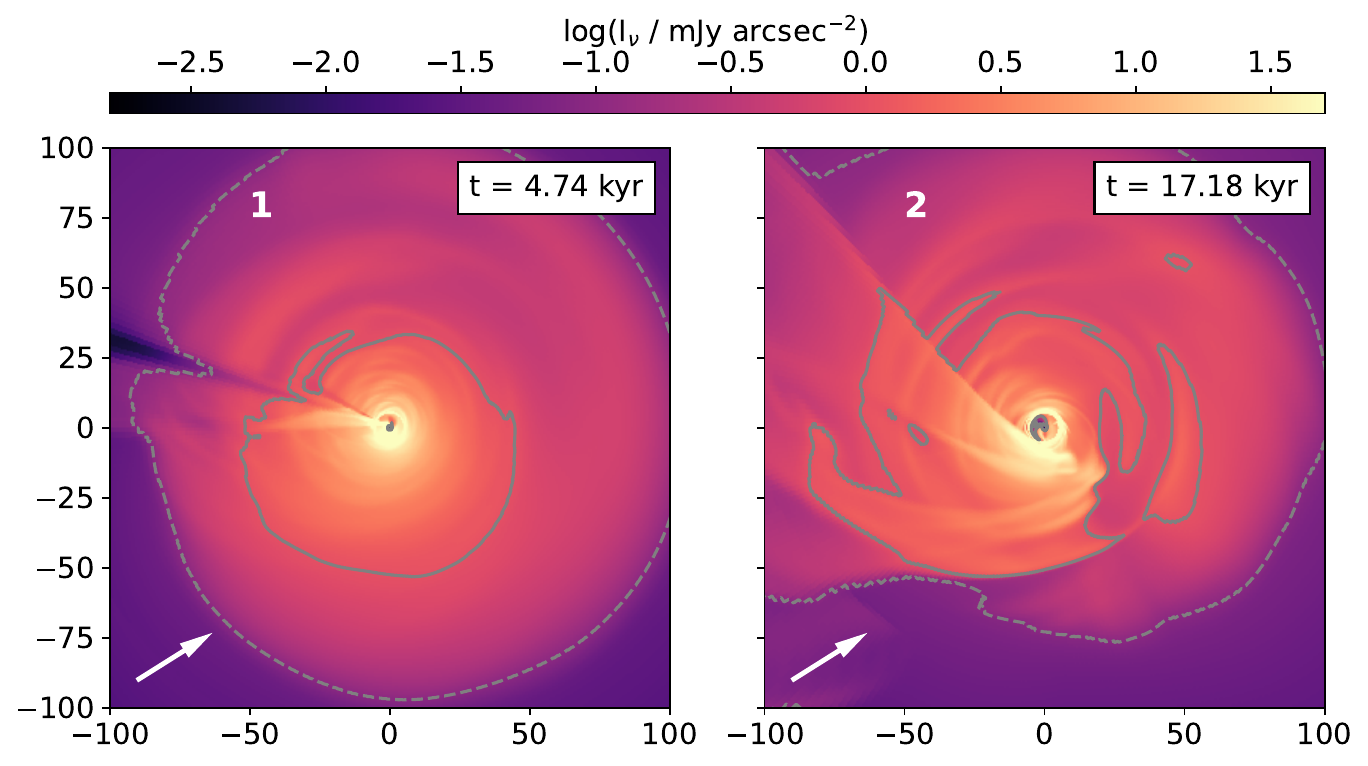}
    \caption{Same as Fig.~\ref{fig:low_spirals_rphi}, but with an artificially high dust-to-gas ratio of 0.1, reproducing the dust mass used for simulations 1 and 2.}
    \label{fig:low_spirals_rphi_highdtg}
\end{figure*}%
\begin{figure*}[htp]
    \centering\includegraphics[width=.8\textwidth]{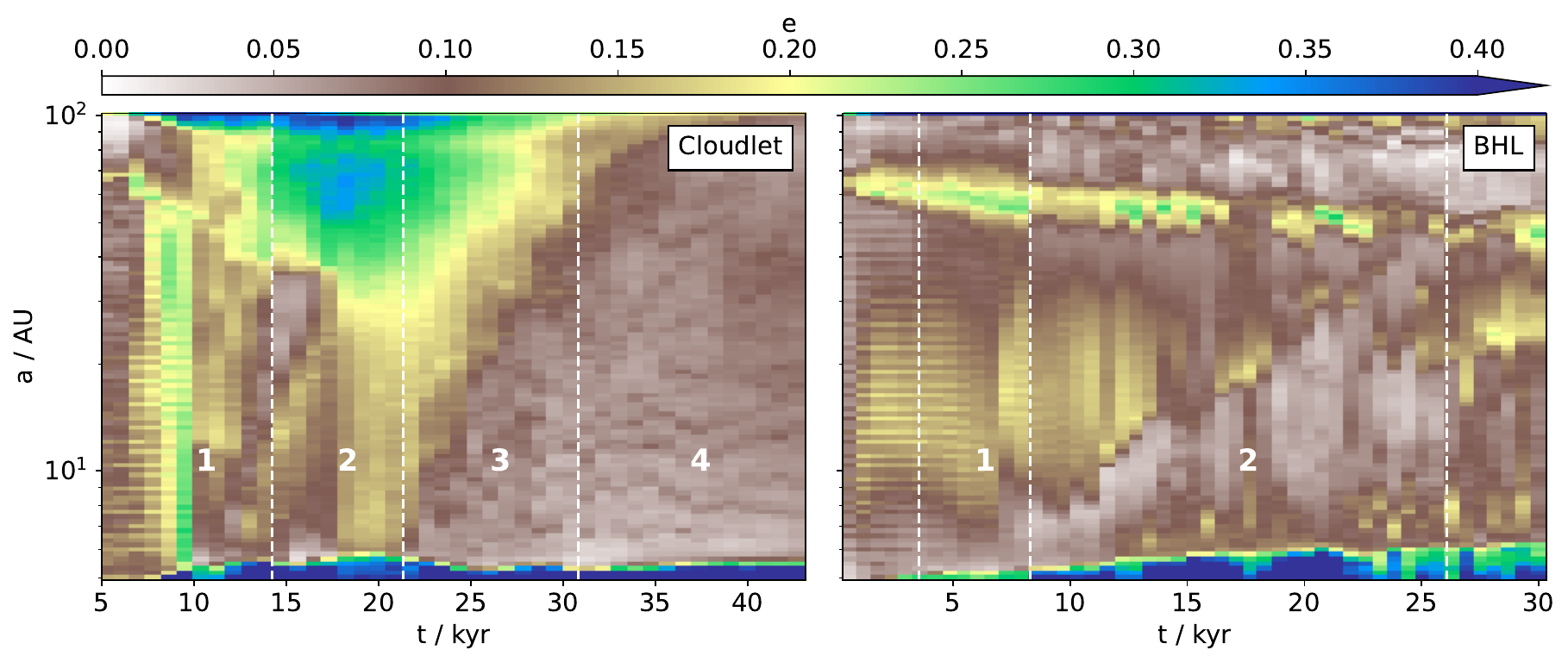}
    \caption{Same as the left panel of Fig.~\ref{fig:cldl_orbit}, but for simulation 3 (left) and simulation 4 (right).}
    \label{fig:low_e}
\end{figure*}%
In this Appendix, we present additional Figures for simulations number 3 and 4, which have a lower disk mass, making them more susceptible to deeper perturbations. In Fig.~\ref{fig:low_spirals_rphi}, we show the scattered light synthetic images of a selection of the snapshots considered for the regular disk mass simulations. The visible structures match the findings presented in Fig.~\ref{fig:comp_spectrum}: For the cloudlet capture, $m=2$ spirals disappear in the inner disk and become weaker in the outer regions, whereas the $m=1$ mode becomes stronger, especially in the outer disk. For the BHL accretion, the $m=2$ mode actually becomes more strongly visible during phase 1, but structures mostly disappear during phase 2. In general, the observed structure is comprised less of flocculent spiral arms. To test to what extent the change in visible structures is related to the change in the scattering height, we also created synthetic scattered light observations with a dust-to-gas ratio higher by a factor of 10 compared to the regular value. Even though this is an unrealistic value, it allows us to reproduce the solid mass assumed in simulations 1 and 2. The results are shown in Fig.~\ref{fig:low_spirals_rphi_highdtg}, showing that the change in emission height does not account for all differences found in the lower disk mass case, but it can account for some changes in visible structure. For the cloudlet case, the spiral structure seems more dominated by the $m=2$ mode for the higher dust-to-gas ratio during phase 2, as is the case for simulation 1, and some spiral structure is recovered in the inner disk. For the BHL accretion case, the $m=2$ mode during phase 1 disappears with the higher dust-to-gas ratio, which also holds true for the flocculent variations during phase 2.

In Fig.~\ref{fig:low_e}, we show the eccentricity of the gas for the cases of lower disk mass. The locations of excitation in time and semi-major axis do not change significantly, though for the cloudlet case, the excitation of eccentricity is stronger. This is an expected result, as the momentum of the infalling material has a greater impact on the total momentum of the gas. For the BHL case, the excitation is at a similar level, or lower at a few points in time, but the inner disk becomes more excited.

\begin{figure*}[htp]
    \centering\includegraphics[width=.45\linewidth]{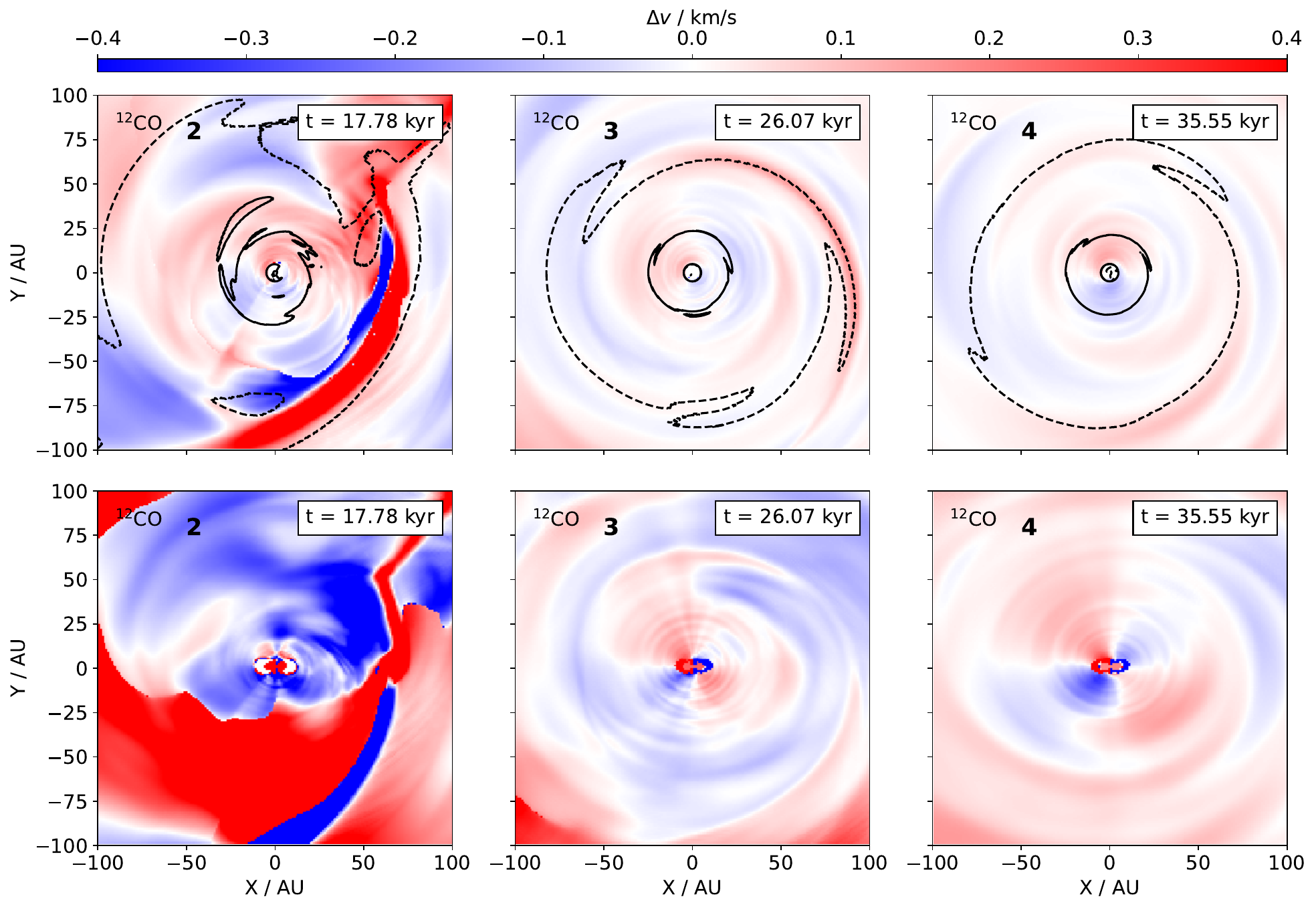}\includegraphics[width=.45\linewidth]{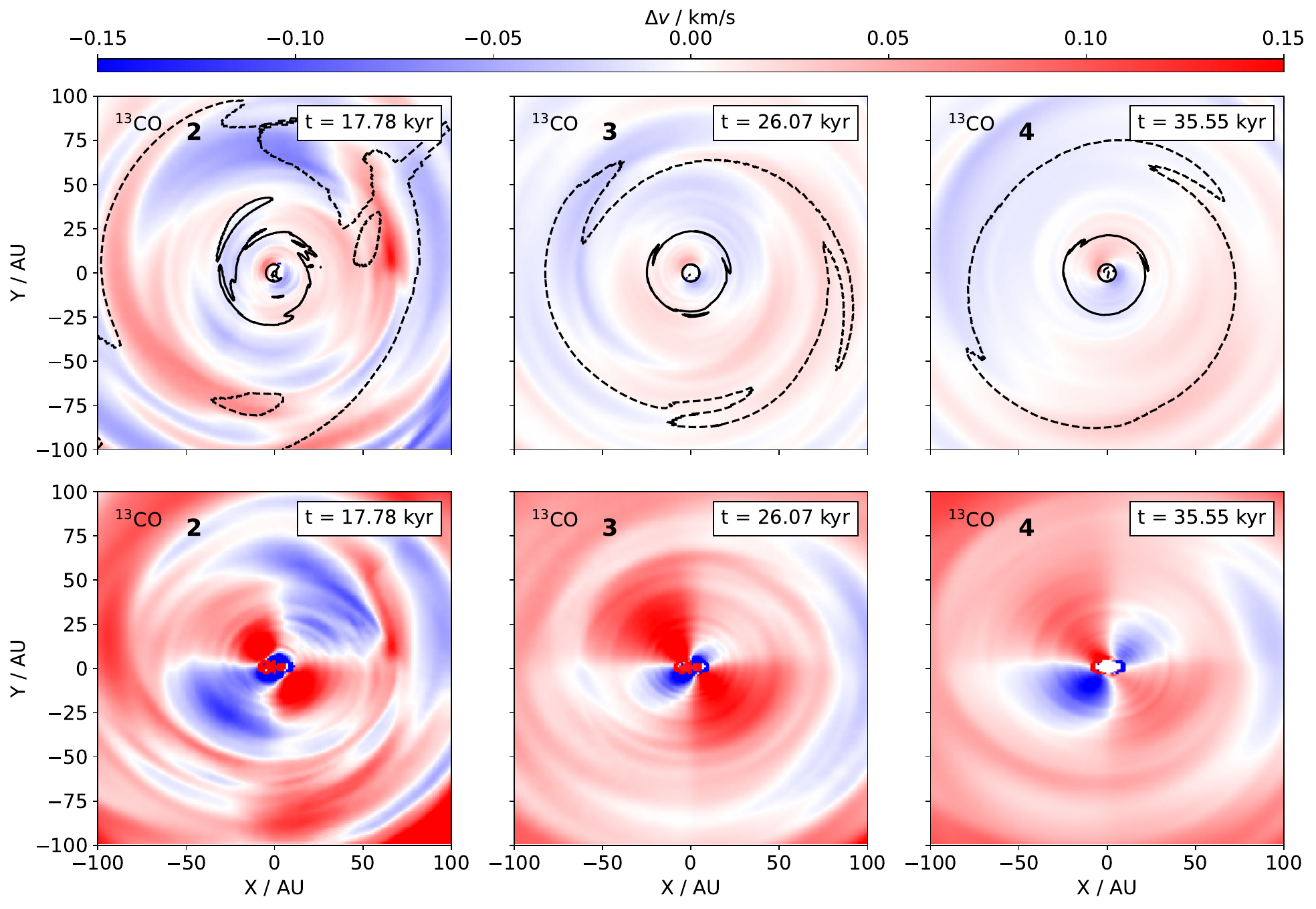}
    \caption{Same as Fig.~\ref{fig:cldl_co} (left) and Fig.~\ref{fig:cldl_13co} (right), but for simulation 3.}
    \label{fig:cldh_xco}
\end{figure*}%
\begin{figure*}[htp]
    \centering\includegraphics[width=.45\linewidth]{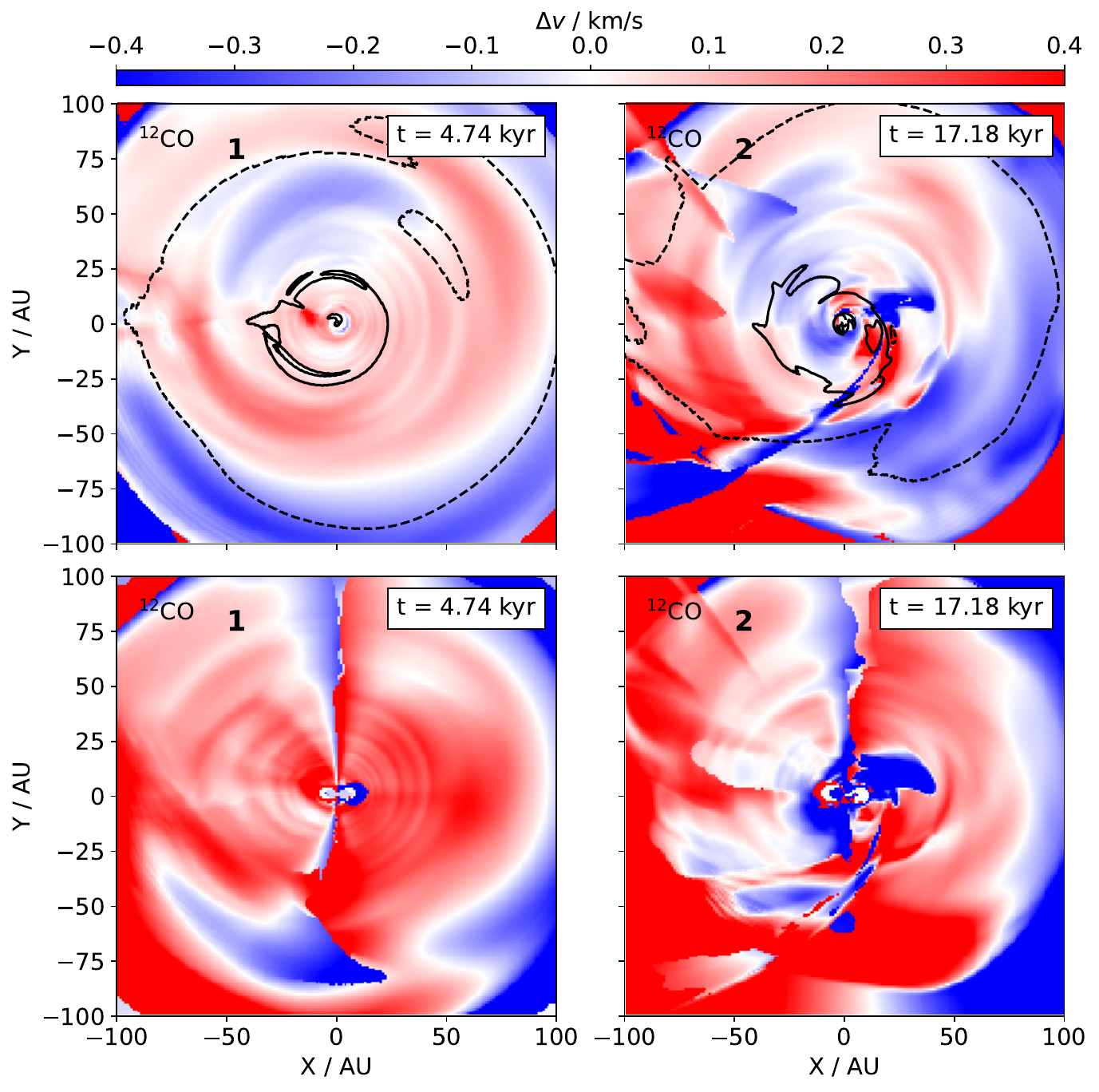}\includegraphics[width=.45\linewidth]{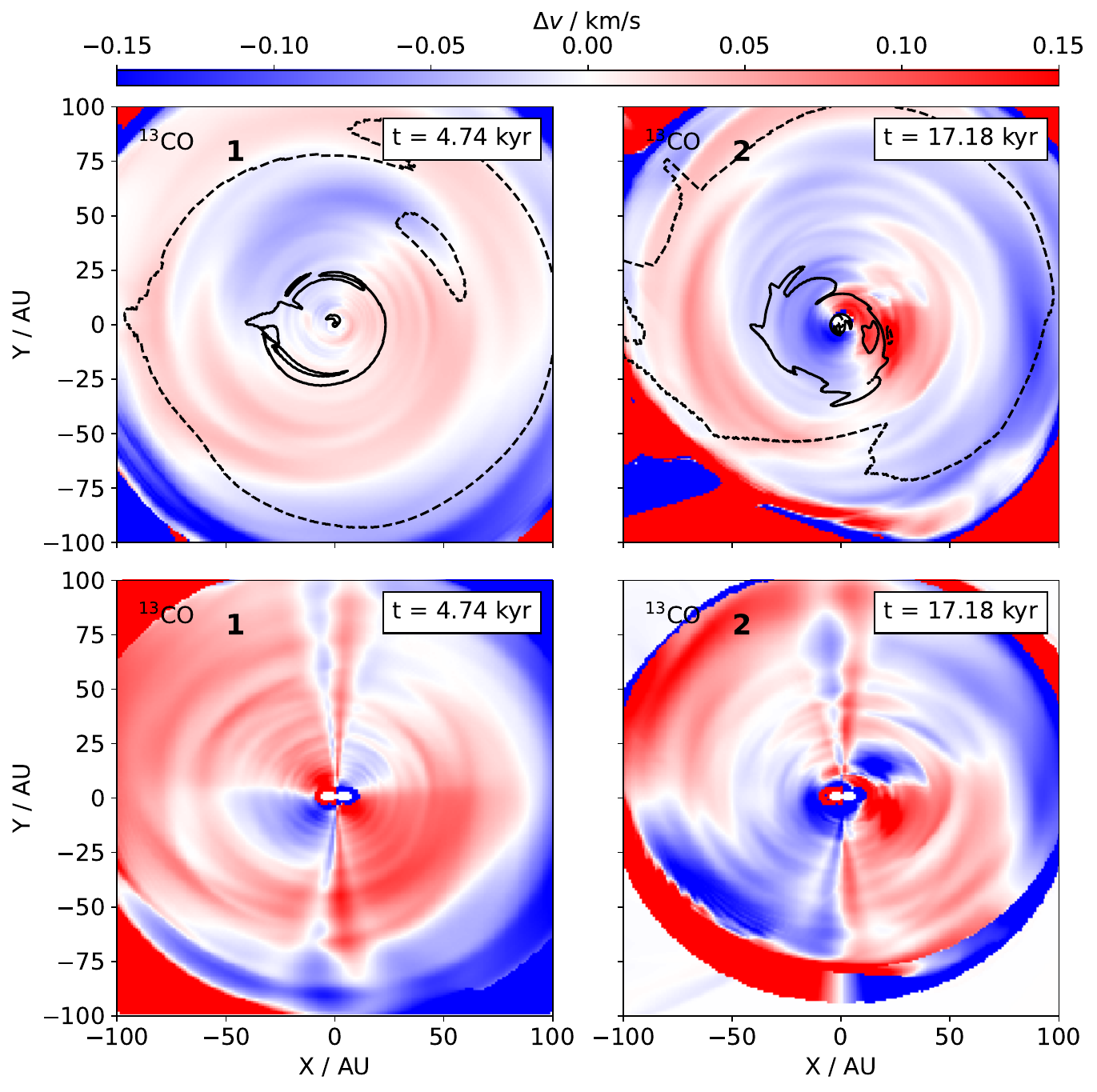}
    \caption{Same as Fig.~\ref{fig:cldl_co} (left) and Fig.~\ref{fig:cldl_13co} (right), but for simulation 4.}
    \label{fig:bhll_xco}
\end{figure*}%
In Fig.~\ref{fig:cldh_xco}, the residuals of the gas motion as determined from the first moment map of the CO isotopologue line emission are shown for the cloudlet capture case. In the face-on images, $m=1$ spiral structure dominates for phases 3 and 4, and they can be seen with more clarity than for the regular disk mass case. For the panel showing the phase 2 snapshot, the structure is less obvious and spirals are more flocculent. Similar to the high disk mass case, the issue of fitting a Keplerian model remains and is further amplified by the stronger perturbations that the disk with lower mass is subject to, creating further severe non-physical residuals in the $i=\SI{30}{\degree}$ images. We show the same kind of images for the BHL accretion case in Fig.~\ref{fig:bhll_xco}, and find an analogously increased prominence of low-mode spirals. Most notably, the $m=1$ spiral structure is very clear in this case, especially during phase 1.

\begin{figure*}[htp]
    \centering\includegraphics[width=.9\linewidth]{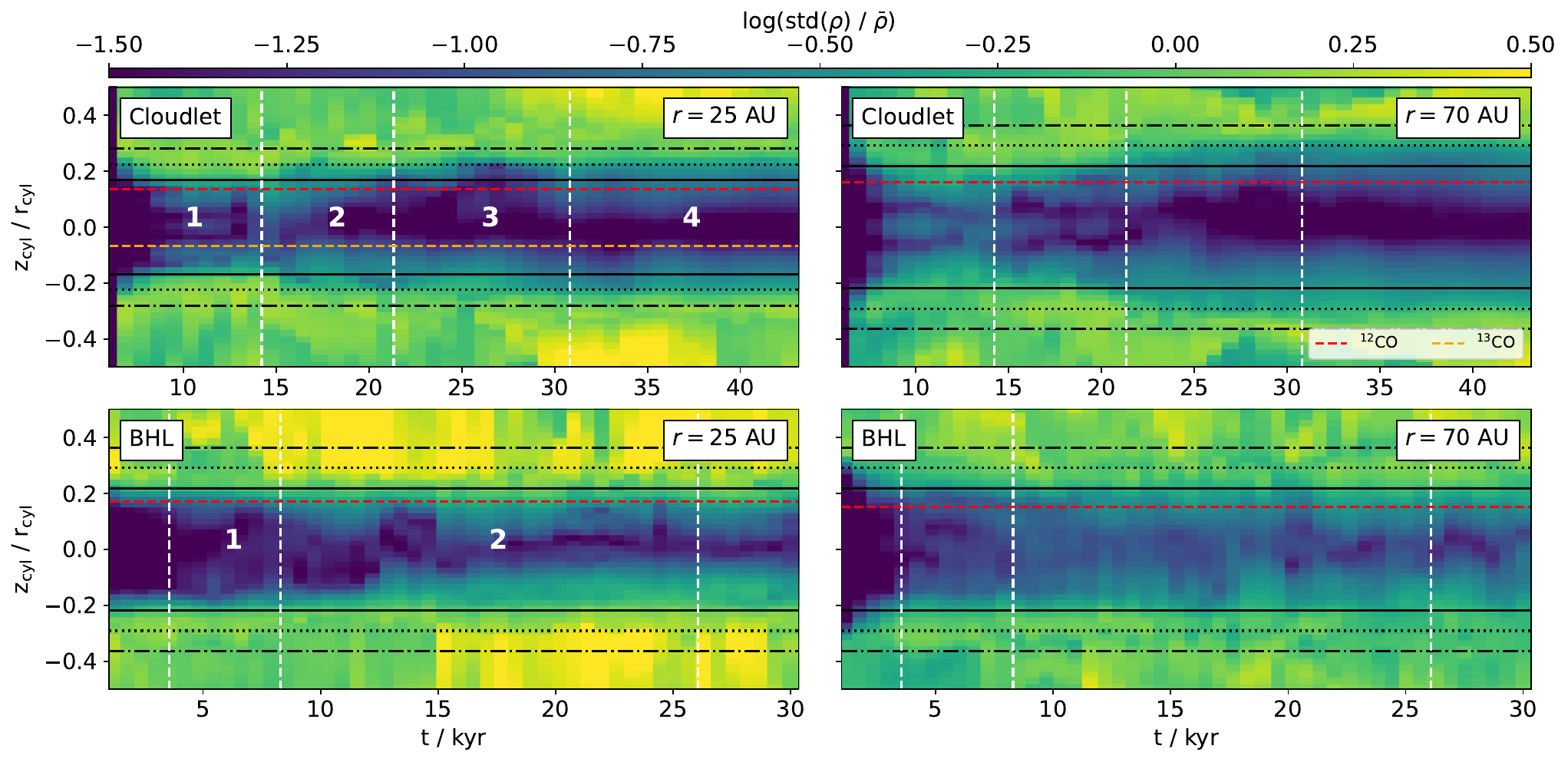}
    \caption{Same as Fig.~\ref{fig:cldl_depth}, but for simulation 3 (top) and simulation 4 (bottom).}
    \label{fig:low_depth}
\end{figure*}%
The depth of the perturbations induced in the lower-mass disks are shown in Fig.~\ref{fig:low_depth}, providing more context on the analysis based on a threshold value shown in Fig.~\ref{fig:comp_depth}. In line with expectations, perturbations protrude further toward the midplane. In the cloudlet case, even the midplane can be weakly affected directly at the time of the main encounter, but they retract to the surface for the later accretion stages, affecting only layers at $z\gtrapprox 3H$. For BHL accretion, deeper layers are affected only at later times, but for significantly longer compared to the cloudlet case. Perturbations occur close to the midplane during the majority of phase 2.
\FloatBarrier

\section{Inclined infall}\label{sec:app_inclined}
\begin{figure*}[htp]
    \centering
    \includegraphics[width=.45\linewidth]{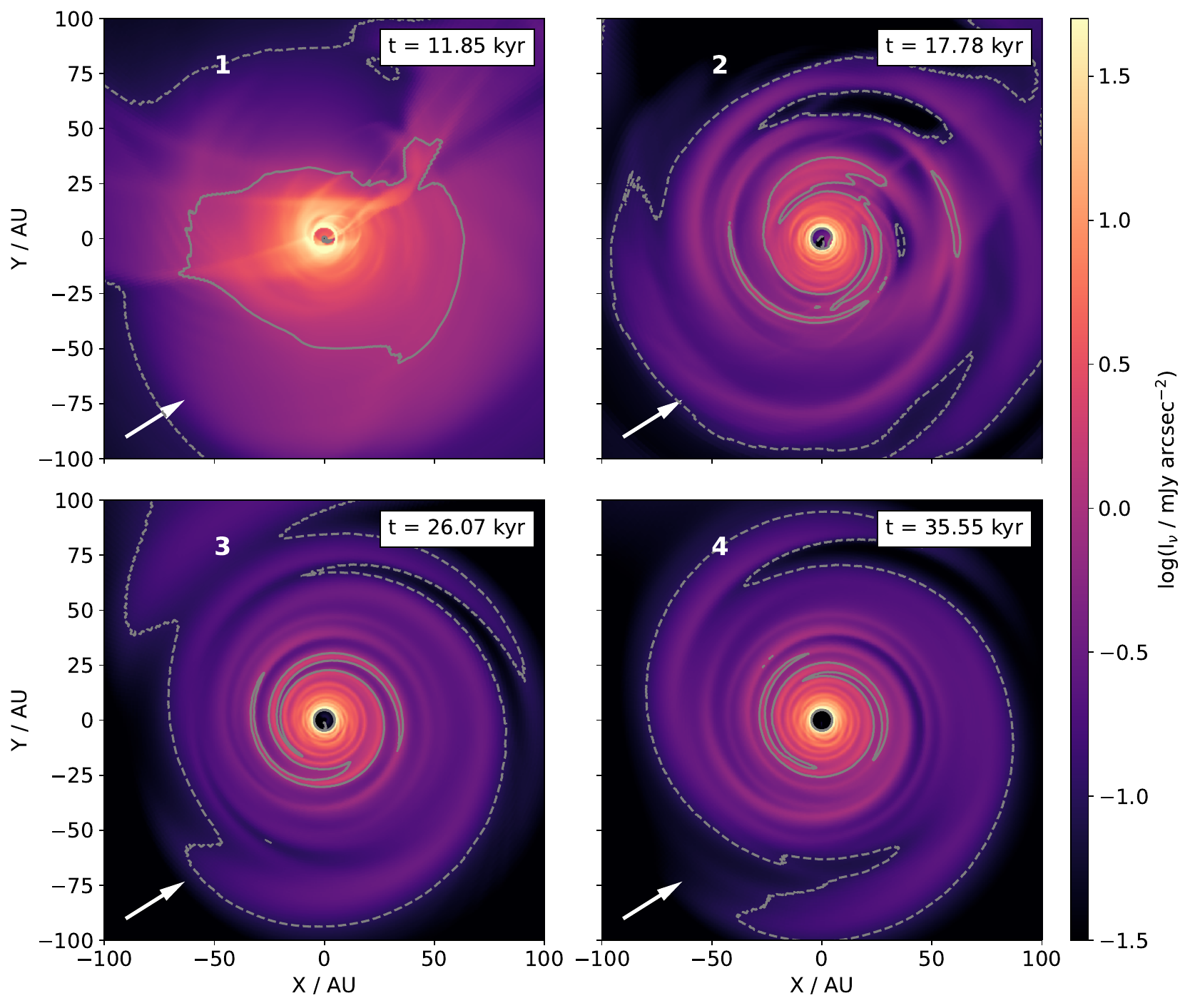}\includegraphics[width=.45\linewidth]{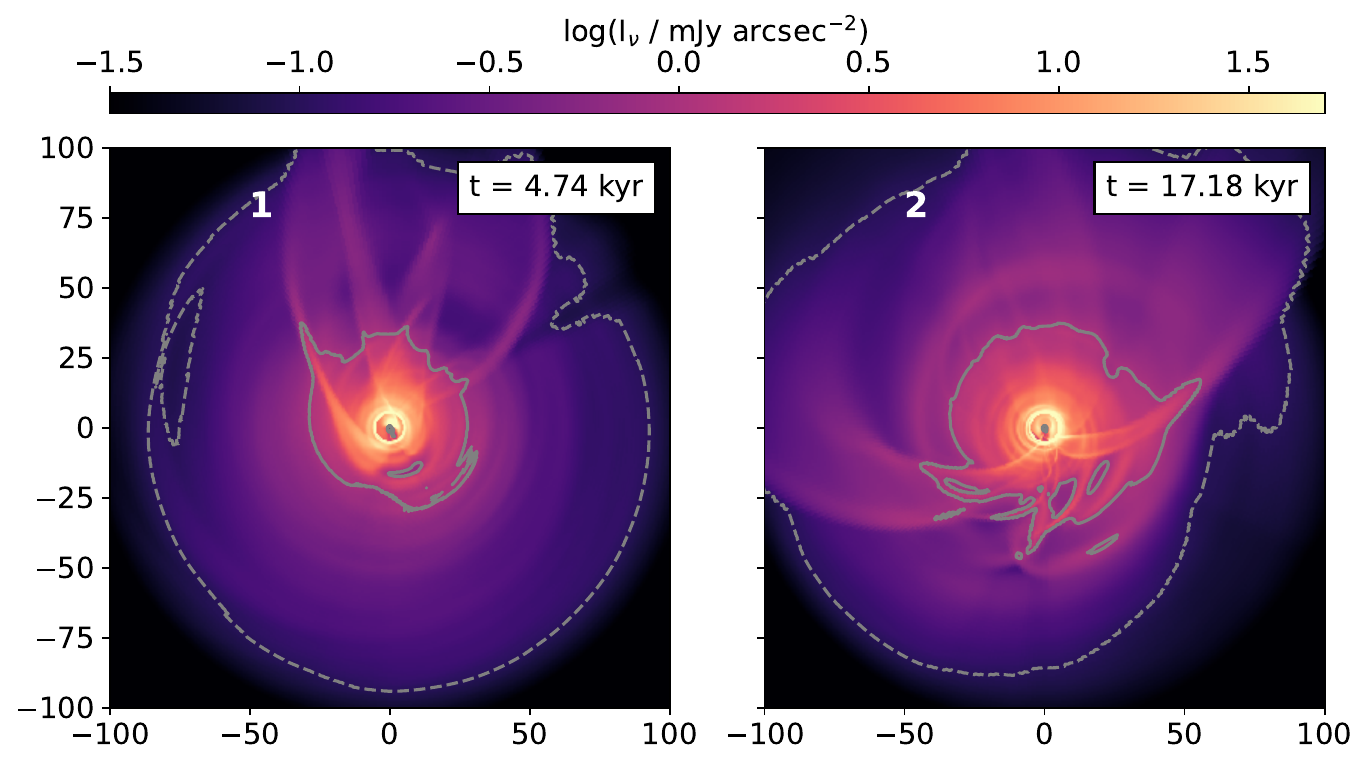}
    \caption{Left: Same as zoomed-in areas of Fig. \ref{fig:cldl_spirals_rphi}. Right: Same as zoomed-in areas of Fig. \ref{fig:bhlh_spirals_rphi}.}
    \label{fig:inclined_spirals_rphi}
\end{figure*}%
In this Appendix, we present synthetic scattered light observations of two simulations where the infall direction is inclined relative to the disk plane. They are shown in Fig. \ref{fig:inclined_spirals_rphi}. The cloudlet capture simulation was run with the same parameters as simulation 1, but the position vector and initial velocity of the cloudlet were rotated by \SI{30}{\degree} around the $x$-axis. The BHL accretion simulation was run with the same parameters as simulation 2, but the systemic velocity unit vector is $\hat{\vec v}_\mathrm{sys}=(0,\frac{1}{\sqrt{2}},\frac{1}{\sqrt{2}})$. In the cloudlet case, $m=2$ spirals arise during the same phases as in the case of in-plane cloudlet capture, but they are less prominent during phase 3. We suspect that this is because the infall perturbations are more symmetrically spread across the disk surface, diminishing the effect. In the limit of perfectly face-on accretion event, we would not expect that spirals form directly. In this case of inclined infall, the disk is also more susceptible to other spiral formation mechanisms, like the excitation of eccentricity \citep{calcino2025a}, or disk warping \citep{winter2025}, which could contribute to the visible spiral patterns. A detailed analysis of this scenario is outside the scope of this work. The accretion tail found in the BHL accretion simulation has an orientation relative to the observer that is unfavorable for the study of spiral patterns, though infall streamers are more easily observed (see HD25).

\FloatBarrier
\section{Warp structure}\label{sec:app_warp}
\begin{figure*}[htp]
    \centering\includegraphics[width=.9\textwidth]{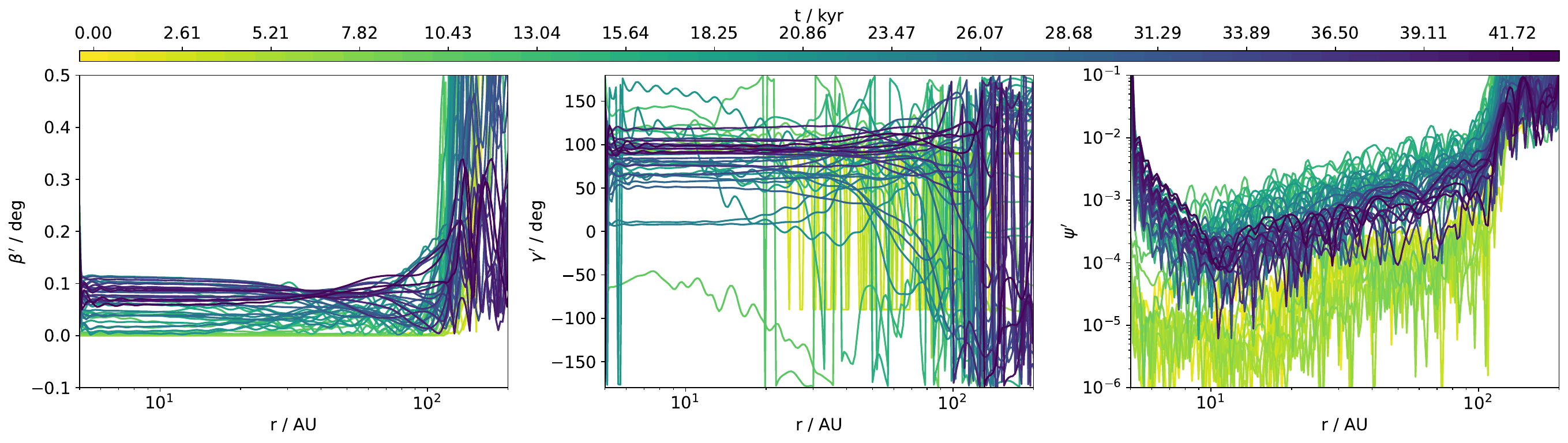}
    \caption{Warp properties as a function of radial distance and time, where a darker color denotes a later snapshot. Left: Disk tilt in the reference frame described by the total angular momentum, $\beta'$. Middle: Disk twist in the same reference frame, $\gamma'$. Right: Warp strength $\psi'$ in that reference frame. All values are for simulation 1.}
    \label{fig:cldl_warp}
\end{figure*}%
\begin{figure*}[htp]
    \centering\includegraphics[width=.9\textwidth]{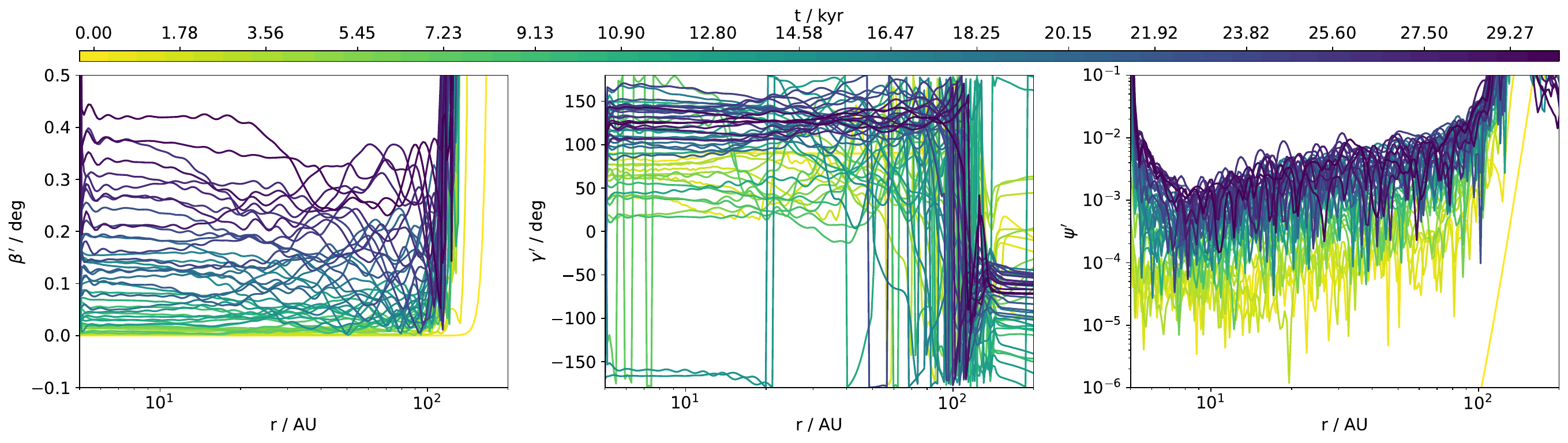}
    \caption{Same as Fig.~\ref{fig:cldl_warp}, but for simulation 2.}
    \label{fig:bhlh_warp}
\end{figure*}%
\begin{figure*}[htp]
    \centering\includegraphics[width=.9\textwidth]{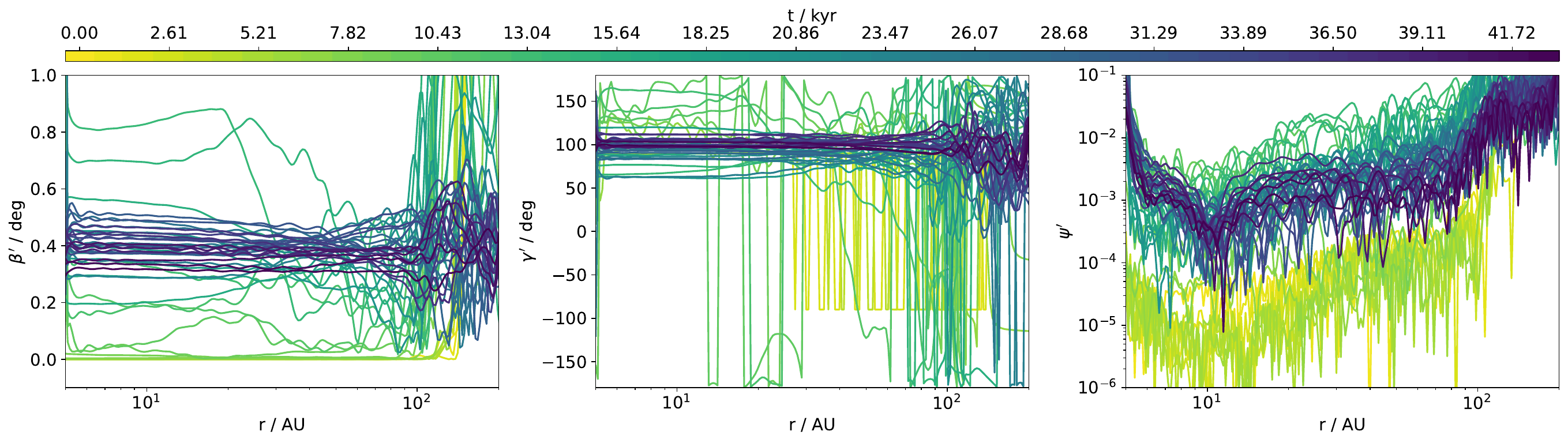}
    \caption{Same as Fig.~\ref{fig:cldl_warp}, but for simulation 3. We note that the range of values on the ordinate is increased compared to that Figure.}
    \label{fig:cldh_warp}
\end{figure*}%
\begin{figure*}[htp]
    \centering\includegraphics[width=.9\textwidth]{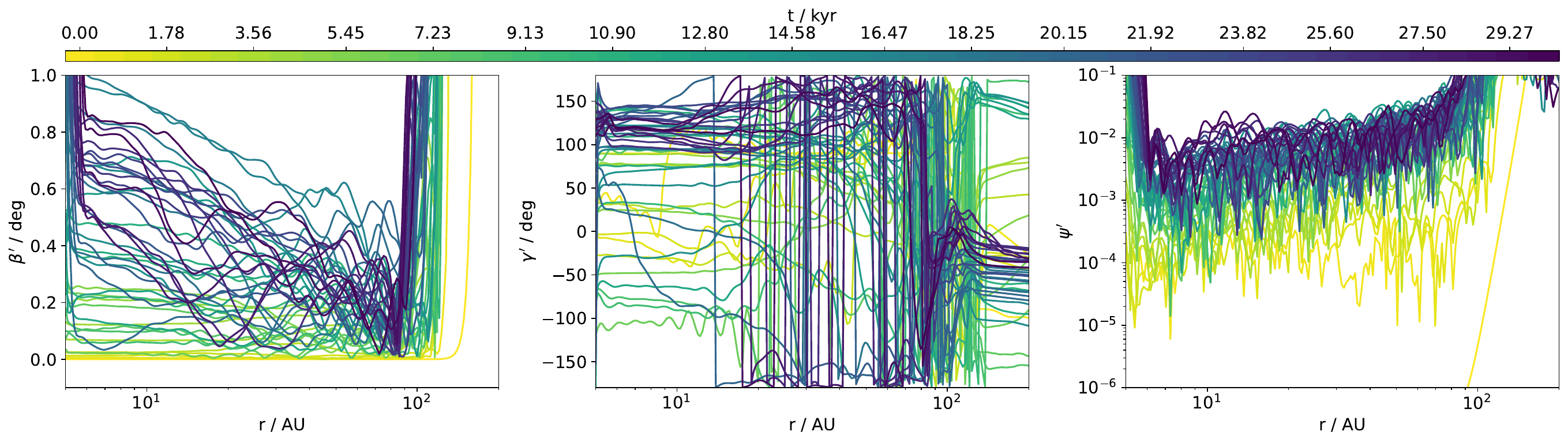}
    \caption{Same as Fig.~\ref{fig:cldl_warp}, but for simulation 4. We note that the range of values on the ordinate is increased compared to that Figure.}
    \label{fig:bhll_warp}
\end{figure*}%
Previous work has shown that a warped and twisted disk can create spiral structures, and that this mechanism can explain aspects of the spiral structure in MWC 758 \citep{winter2025}. While our simulation setups was not chosen to promote the formation of a disk warp, we nevertheless investigated whether a warp could be the cause of the substructures we find in this Appendix. Figures \ref{fig:cldl_warp},\ref{fig:bhlh_warp},\ref{fig:cldh_warp} and \ref{fig:bhll_warp} show three parameters that are commonly considered for the characterization of a warped disk structure for all four simulations. These parameters are the disk tilt $\beta$,
\begin{equation}
\beta(r)=\arccos(l_z(r))\text{,}
\end{equation}
the twist $\gamma$,
\begin{equation}
\gamma(r)=\arctan_2(l_y(r),l_x(r))\text{,}
\end{equation}
and the warp strength $\psi$,
\begin{equation}
\psi(r)=\left|\vec\psi(r)\right|=r\left|\frac{\partial\vec l(r)}{\partial r}\right|\text{.}
\end{equation}
Here, $\vec l$ is the unit vector of the cell volume-weighted average angular momentum in a spherical shell at radius $r$ with a thickness given by the grid resolution, $\Delta r$. They are calculated in a different reference frame than that one given by our grid configuration; it is defined as the frame where the total disk angular momentum coincides with the $z$ direction, so that the transformation is given by
\begin{equation}
    \vec l' = R_x(-\phi_x)R_z(-\phi_z)\vec l\text{,}
\end{equation}
where $R_i$ are rotation matrices about axis $i$, and the rotation angles are given by
\begin{align}
    \phi_x &= \arccos(l_\mathrm{tot,z}),\\
    \phi_z &= \arccos\left(-\frac{l_\mathrm{tot,y}}{\sqrt{1-l_\mathrm{tot,z}^2}}\right),
\end{align}
with $\vec l_\mathrm{tot}$ being the unit vector of the total disk angular momentum, integrated within \SI{150}{\astronomicalunit}.
The initial condition of the cloudlet capture simulations lead to a capture in the plane of the disk. We therefore do not expect a strong warp to occur, which we confirm with this analysis. In the regular disk mass case, the maximal tilt in the bulk of the disk is \SI{0.1}{\degree}. Here, there is no considerable twist, that is, $\gamma$ does not change significantly with radius, and the maximal strength is $\psi\sim\num{e-2}$. The warp is not much more significant for the lower disk mass case, where the maximal tilt is \SI{0.8}{\degree}, there is no twist, and the strength only reaches $\psi\sim\num{e-1}$ in the outer disk.

For the simulations of BHL accretion, the systemic velocity is also in the disk plane, which does not promote warping of the disk, but the turbulent nature of the environment could, in principle, lead to more significant warping. However, we also do not find this to be the case. For the regular disk mass case, the tilt reaches $\beta\sim\SI{0.4}{\degree}$ and the strength reaches $\psi\sim\num{e-2}$. For the low disk mass case, the strength is of the same order of magnitude, but lasts longer, and the maximum tilt reaches close to $\beta\sim\SI{1}{\degree}$. Neither of these simulations exhibit considerable twisting. We therefore conclude that warping is not the source of our CO line emission substructures, as expected. It remains to be investigated how out-of-plane accretion could cause more significant warp effects, which was found in previous works about late infall (e.g., \citealt{kuffmeier2021}), and how that might cause the creation of substructures.
\FloatBarrier\newpage
\section{Discminer fit parameters}\label{sec:app_discminer}
\begin{table*}[htp]
    \centering
    \caption{Model parameters determined by the \texttt{discminer} code for simulation 1, phase 2.}
    \begin{tabular}{llcccc}
        \hline
        \hline
        Attribute & Parameter & $^{12}$CO face-on & $^{13}$CO face-on & $^{12}$CO inclined & $^{13}$CO inclined\\
        \hline
        Orientation & $i$ (rad) & -0.013918 & -0.000893 & 0.531263 & 0.534741\\
         & PA (rad) & 2.473116 & 1.777206 & 3.078874 & 3.13764\\
        \hline
        Peak intensity & $I_0$ (Jy/pix) & 1.786355 & 1.636759 & 3.400697 & 3.756422\\
         & $p$ & -1.756626 & -1.358126 & -3.386501 & -1.62514\\
         & $q$ & 0.853544 & 0.914558 & 1.464581 & 1.15346\\
         & $R_\mathrm{out}$ (AU) & 292.028464 & 173.065857 & 337.543622 & 295.388881\\
        \hline
        Line width & $L_0$ (km/s) & 0.443715 & 0.360182 & 0.182926 & 0.215174\\
         & $p$ & 1.437235 & -0.591546 & 3.854614 & -0.155465\\
         & $q$ & -0.656888 & 0.347203 & -1.791307 & -0.099281\\
        \hline
        Line slope & $L_s$ & 0.763965 & 3.133948 & 1.187412 & 2.666266\\
         & $p$ & -0.547374 & -0.193224 & 0.18797 & 0.14124\\
        \hline
        Upper surface & $z_0$ (AU) & 75.931583 & 24.582231 & 92.94603 & 44.740208\\
         & $p$ & 2.62048 & 1.173912 & 2.451284 & 1.578596\\
         & $R_b$ (AU) & 136.363943 & 157.124266 & 135.436658 & 123.259567\\
         & $q$ & 3.548277 & 3.837051 & 1.179266 & 1.62922\\
        \hline
        Lower surface & $z_0$ (AU) & 10.186404 & 27.280565 & 46.835506 & 21.554529\\
         & $p$ & 1.401177 & 1.738811 & 2.769326 & 1.524081\\
         & $R_b$ (AU) & 156.521556 & 147.455423 & 191.508056 & 153.658222\\
         & $q$ & 4.44322 & 4.507593 & 1.997903 & 2.751715\\
        \hline
    \end{tabular}
    \label{tab:discminer_1_2}
\end{table*}%
\begin{table*}[htp]
    \centering
    \caption{Model parameters determined by the \texttt{discminer} code for simulation 1, phase 3.}
    \begin{tabular}{llcccc}
        \hline
        \hline
        Attribute & Parameter & $^{12}$CO face-on & $^{13}$CO face-on & $^{12}$CO inclined & $^{13}$CO inclined\\
        \hline
        Orientation & $i$ (rad) & -0.008464 & 0.000284 & 0.541732 & 0.53171\\
         & PA (rad) & 1.236967 & 1.665269 & 3.137841 & 3.139437\\
        \hline
        Peak intensity & $I_0$ (Jy/pix) & 2.181766 & 1.501563 & 4.827026 & 4.737016\\
         & $p$ & -0.47821 & -0.859487 & -1.742632 & -1.666194\\
         & $q$ & 0.918481 & 0.551083 & 1.020335 & 1.179157\\
         & $R_\mathrm{out}$ (AU) & 213.873871 & 215.980399 & 252.403358 & 320.514226\\
        \hline
        Line width & $L_0$ (km/s) & 0.333042 & 0.446096 & 0.208635 & 0.17185\\
         & $p$ & -0.202836 & -0.671717 & 0.366968 & -0.162427\\
         & $q$ & 0.091417 & 0.550841 & -0.400435 & -0.142254\\
        \hline
        Line slope & $L_s$ & 2.652136 & 3.471416 & 1.236261 & 3.530185\\
         & $p$ & -0.185746 & -0.096082 & -0.743685 & 0.182489\\
        \hline
        Upper surface & $z_0$ (AU) & 44.314921 & 20.546271 & 43.099641 & 43.91231\\
         & $p$ & 0.662675 & 1.502242 & 1.79254 & 1.687935\\
         & $R_b$ (AU) & 141.891144 & 101.516024 & 136.948374 & 114.09979\\
         & $q$ & 6.003319 & 2.704703 & 3.515581 & 1.596887\\
        \hline
        Lower surface & $z_0$ (AU) & 74.961395 & 31.079452 & 16.12589 & 26.154063\\
         & $p$ & 0.417348 & 1.031362 & 1.820823 & 1.727622\\
         & $R_b$ (AU) & 136.507681 & 143.967677 & 317.521125 & 125.081057\\
         & $q$ & 6.400883 & 4.939914 & 3.330637 & 1.55085\\
        \hline
    \end{tabular}
    \label{tab:discminer_1_3}
\end{table*}%
\begin{table*}[htp]
    \centering
    \caption{Model parameters determined by the \texttt{discminer} code for simulation 1, phase 4.}
    \begin{tabular}{llcccc}
        \hline
        \hline
        Attribute & Parameter & $^{12}$CO face-on & $^{13}$CO face-on & $^{12}$CO inclined & $^{13}$CO inclined\\
        \hline
        Orientation & $i$ (rad) & 0.003787 & -0.000168 & 0.54285 & 0.530572\\
         & PA (rad) & 3.000773 & 0.924257 & 3.138717 & 3.13899\\
        \hline
        Peak intensity & $I_0$ (Jy/pix) & 2.16589 & 1.842631 & 4.249739 & 4.764129\\
         & $p$ & -0.431873 & -1.113991 & -0.962164 & -1.408765\\
         & $q$ & 1.02278 & 0.753264 & 0.730904 & 1.102828\\
         & $R_\mathrm{out}$ (AU) & 207.641656 & 184.231506 & 175.519168 & 344.843874\\
        \hline
        Line width & $L_0$ (km/s) & 0.326406 & 0.52853 & 0.213381 & 0.168953\\
         & $p$ & -0.233479 & -0.908925 & 0.063096 & -0.176727\\
         & $q$ & 0.28306 & 0.771579 & -0.264783 & -0.13784\\
        \hline
        Line slope & $L_s$ & 3.329383 & 3.498086 & 2.42813 & 3.718619\\
         & $p$ & -0.058892 & -0.07426 & -0.097767 & 0.225348\\
        \hline
        Upper surface & $z_0$ (AU) & 60.525584 & 11.221921 & 28.781314 & 41.804518\\
         & $p$ & 0.360929 & 1.765037 & 1.178712 & 1.5583\\
         & $R_b$ (AU) & 143.208523 & 222.033692 & 128.884887 & 108.790378\\
         & $q$ & 6.104135 & 3.285206 & 7.933757 & 1.547284\\
        \hline
        Lower surface & $z_0$ (AU) & 62.817536 & 33.433312 & 15.393315 & 20.914966\\
         & $p$ & 0.312927 & 0.952737 & 1.535566 & 1.553152\\
         & $R_b$ (AU) & 147.570851 & 155.267855 & 120.549235 & 128.287544\\
         & $q$ & 6.952082 & 4.70432 & 0.7615 & 1.926948\\
        \hline
    \end{tabular}
    \label{tab:discminer_1_4}
\end{table*}%
\begin{table*}[htp]
    \centering
    \caption{Model parameters determined by the \texttt{discminer} code for simulation 2.}
    \begin{tabular}{llcc}
        \hline
        \hline
        Attribute & Parameter & $^{12}$CO phase 1 & $^{12}$CO phase 2\\
        \hline
        Orientation & $i$ (rad) & 0.035227 & 0.015317\\
         & PA (rad) & 3.031075 & 2.786731\\
        \hline
        Peak intensity & $I_0$ (Jy/pix) & 4.019554 & 1.659029\\
         & $p$ & -0.553326 & -4.422599\\
         & $q$ & 1.489104 & 1.781956\\
         & $R_\mathrm{out}$ (AU) & 289.739617 & 198.899374\\
        \hline
        Line width & $L_0$ (km/s) & 0.299288 & 0.386206\\
         & $p$ & -0.083432 & 3.326358\\
         & $q$ & -0.437694 & -1.387461\\
        \hline
        Line slope & $L_s$ & 1.745675 & 1.649106\\
         & $p$ & -0.142413 & -0.073902\\
        \hline
        Upper surface & $z_0$ (AU) & 55.022357 & 96.445318\\
         & $p$ & 0.262799 & 2.564996\\
         & $R_b$ (AU) & 130.580834 & 130.327168\\
         & $q$ & 9.019506 & 8.866092\\
        \hline
        Lower surface & $z_0$ (AU) & 28.880108 & 90.17602\\
         & $p$ & 1.2896 & 3.445089\\
         & $R_b$ (AU) & 111.040741 & 126.227747\\
         & $q$ & 4.293829 & 9.086012\\
        \hline
    \end{tabular}
    \label{tab:discminer_2}
\end{table*}%
\end{appendix}
\end{document}